\documentclass[preprint,aps]{revtex4}

\usepackage[final]{graphicx}
\usepackage{dcolumn}
\usepackage{bm}
\usepackage{braket}
\usepackage{amsmath}
\usepackage{graphicx}
\usepackage{float}
\begin{document}

\title{Cavity QED with tunneling: an application of adiabatic elimination in quantum trajectory theory}
\author{Charlie Baldwin}
\author{Perry Rice}
\affiliation{Macklin Quantum Information Sciences\\Department of Physics, Miami University, Oxford OH 45056}

\begin{abstract}
We model the evolution of a single atom in a cavity interacting with two lasers, one far off resonance which creates an optical potential lattice and one near resonance, which can interact with the atom.  The atom may tunnel between sites in the lattice.  We find the photon counting statistic $g^{(2)}(\tau)$ depends on the tunneling rate.  Additionally we derive methods to work with a Bose-Einstein Condensate in the same situation, in ther Josephson junction regime.  It is shown that oscillations in $g^{(2)}(\tau)$ are directly proportional to the tunneling coefficient. We work in the bad-cavity limit and give an example of the use of adiabatic elimination in quantum trajectory theory
\end{abstract}

\maketitle

\setcounter{page}{1}

\section{Introduction}
Cavity quantum electrodynamics (cQED) is the study of ultra-cold atoms interacting with laser light in a high quality mirror cavity.

In this paper, we model a single atom in an optical lattice, within a cavity, allowed to tunnel between lattice sites.  The model provides a base for further work with multiple atoms, which will be able to interact.  We treat the atoms as a Bose-Einstein condensate which will be discussed later. We establish methods to study the particle's behavior by observing emission from the cavity.

There have been many variations of the cQED system studied.  We will explain the basic model here.  We place the atoms inside a high finesse mirror cavity.  The atoms are assumed to only have two allowed energy levels, excited ($\ket{e}$) and ground ($\ket{g}$).  The cavity is then pumped with a single mode driving laser of strength $Y$ that interacts with the atoms with coupling rate $g$.  The atoms will then oscillate between its two energy levels with the $Rabi$ $frequency$ equal to $g \sqrt{N}$ for $N$ atoms.

This alone, however, is not a full description of the system.  We have failed to include the loss of photons from the cavity.  There are two types of photon loss, fluorescence and transmission.  Fluorescence is a result of the atom absorbing a photon from the driving laser and later emitting it out the side of the cavity in a random direction (spontaneous emission.)  
 In addition to these interactions we add, in this paper, add an optical lattice and tunneling.  The lattice is created by pumping a far off resonant laser into the cavity creating a finite number of sites that the atom can occupy.  The atoms can jump between wells determined by a constant $T$.  We can also treat the atoms as  Bose-Einstein condensate, which requires accounting for additional interactions between atoms.

\begin{center}
\begin{figure}[H] \label{fig: gen setup}
\includegraphics{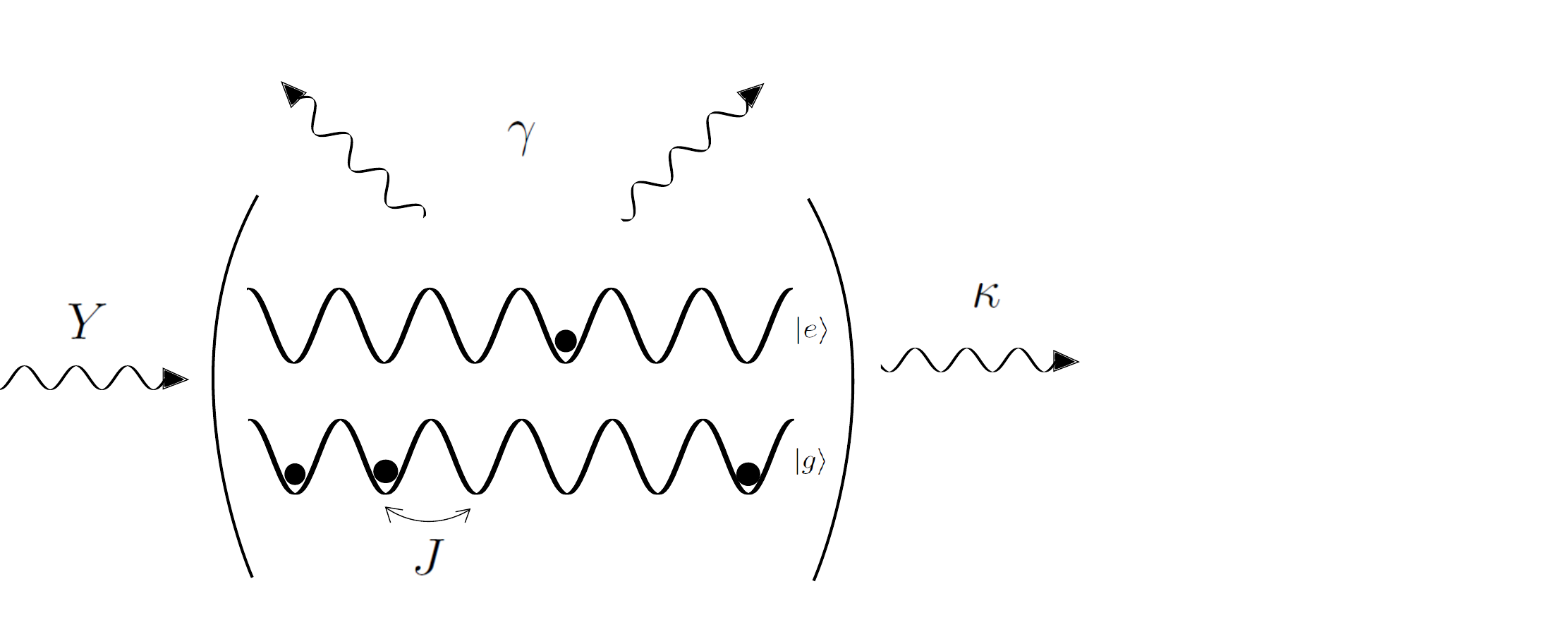}
\caption[General model]{Mirror cavity with incoming laser $Y$, multiple site optical cavity with tunneling constant $J$, spontaneous emission $\gamma$ and cavity loss $\kappa$}
\end{figure}
\end{center}

To solve for the evolution of the system we utilize quantum trajectory theory. A non-Hermitian Hamiltonian accounts for all of these interactions modeling the atoms evolution in the cavity.  Quantum jumps occur which are random events modeling the fluorescence and transmission discussed earlier.  Using times when photons are fluoresced (spontaneous emission) and transmitted (cavity loss.)  we can calculate the photon counting statistics.

In section two, we will outline the system in question.  In section three, we will derive the weak field trajectory model which treats atoms in a cavity interacting with a driving field  and an off resonance lattice beam.  In section four, we discuss the manner in which one uses adiabatic elimnation of variables.  I In section five we will combine the weak field trajectory model and the Bose-Einstein condensate model to give a single Hamiltonian to model $N$ two level boson atoms in an optical lattice interacting with a driving field.  In section six, we will present the numerical model of a single two level atom in a two site optical lattice and compare it to theoretical predictions.  And finally, in section seven we conclude the analysis and give ideas for future studies.

\section{ Optical Lattices} \label{ss: OL}
Eventually our system will include a 1D optical lattice that will restrict the motion of the atoms inside the cavity.  Optical lattices represent a periodic potential created by a set of interfering laser beams.  The potential is caused by the ac-stark shift on neutral atoms \cite{Grondalski}.  The result is the atoms are evenly distributed like a crystal  \cite{Grondalski}.  However, unlike crystals they are easy to manipulate since the size and strength of the lattice sites are determined by the laser creating the lattice. 

The model we use here is semi-classical where we treat atom quantum mechanically but the field classically.  The Hamiltonian then contains two terms, one for the atom and the other for the atom's dipole interaction with the field
\begin{equation}
\hat{H} = \frac{\hbar \omega_0}{2} \hat{\sigma}_z - \vec{E}(z, t) \cdot \hat{d}
\end{equation}
The electric field from a laser is $\vec{E}(z, t) = E_0(z) cos(\omega t) \hat{x}$ oscillates in $\hat{x}$ and polarized in $\hat{x}$, which makes the Hamiltonian
\begin{equation}
\hat{H} = \frac{\hbar \omega_0}{2} \hat{\sigma}_z - \hbar \Omega(z) cos(\omega t) \hat{\sigma}_x
\end{equation}
where $\Omega = E_0(z) \braket{e | \hat{x} \cdot \hat{d} | g}$ and $\hat{\sigma}_x = \hat{\sigma}_+ + \hat{\sigma}_-$ which results from the dot product. The Hamiltonian can be rewritten with the rotating wave approximation, which neglects rapidly oscillating terms
\begin{equation}
\hat{H} \approx \frac{\hbar \omega_0}{2} \hat{\sigma}_z - \frac{\hbar \Omega(z)}{2} \left( e^{-i \omega t} \hat{\sigma}_+ + e^{i \omega t} \hat{\sigma}_- \right)
\end{equation}eq: 2comp BEC Ham
From here we make the transformation to a rotating frame at rate of the laser $\omega$.
\begin{equation}
\hat{H} \approx  -\frac{\hbar\Delta}{2} \hat{\sigma}_z - \frac{\hbar \Omega(z)}{2}\hat{\sigma}_x
\end{equation}
Where $\Delta = \omega - \omega_0$ is the detuning.  The eigenvalues for the Hamiltonian are the energy levels of the system
\begin{equation}
E_{\pm} = \pm \sqrt{\left(\frac{\hbar \Delta}{2} \right)^2  + \left( \frac{\hbar \Omega(z)}{2} \right)^2} \approx \pm \left( \frac{\hbar \Delta}{2} + \frac{\hbar \Omega^2(z)}{4 \Delta} \right)
\end{equation}
This is the energy of a two level atom either in the exited ($+$) or ground state ($-$).  The first term is constant so we can shift the zero point energy to that value.  The second term varies with position.  Since the lattice is formed by a laser we assume it has the form $\Omega(z) = \Omega_0 sin(kz)$ which makes the second term proportional to
\begin{equation}
E_{ls} \propto sin^2(kz)
\end{equation}
where the subscript denotes that this is the light shift energy \cite{Grondalski}.  The light shift is proportional to the potential the atom feels in the optical lattice.  We stated before that the lattice is periodic and this derivation verifies that claim.  The laser's wavelength and intensity define the site spacing and depth respectively.  For our model, we treat the lattice like an external potential.
\begin{equation}
V(z) = V_0 \mbox{sin}^2(kz)
\end{equation}
All of the atoms within the cavity feel an external force dependent on their position.  Later we will show how this external potential affects the behavior of a BEC.

\section{ State Amplitudes and Weak Field Limit} \label{ss: 2comp}
In this section, we will combine all of our analysis thus far to find the state amplitudes of the system.  We will put a limit to the number of photons in the system, known as the weak field limit, in order to keep the basis a manageable size.  The analysis here follows \cite{OptComm} and \cite{CAVQEDBIG}.  
\begin{equation} \label{eq: 2comp Ham}
\hat{H} =  i \frac{\hbar \Delta_a}{2} \hat{J}_z + \hbar \Delta_c \hat{a}^{\dagger} \hat{a} +  i \hbar g (  \hat{a}^{\dagger} \hat{J}_- - \hat{a} \hat{J}_+ ) + i \hbar Y ( \hat{a}^{\dagger} -  \hat{a})- i\hbar (\frac{\gamma}{2} \hat{J}_+ \hat{J}_- + \kappa \hat{a}^{\dagger} \hat{a})
\end{equation}
Here $\Delta_a = \omega - \omega_0$ and $\Delta_c = \omega - \omega_c$ are the detuning between the atom's internal frequency ($\omega_a$) and the cavity mode ($\omega_c$) and the driving frequency ($\omega$) respectfully \cite{OptComm}.   We add the fourth term to represent a driving laser off resonance.  The $\hat{\sigma}$ have also been replaced with $\hat{J}$ operators which are generalized for multiple atoms.  
\begin{equation}
\hat{J}_{\pm} = \sum_{i = 1}^N \hat{\sigma}_{\pm}^i
\end{equation}
Using $\hat{J}$ allows us to look at systems with $N$ atoms which is part of our ultimate goal since dealing with a BEC requires having multiple atoms.  They act on the symmetric states

\begin{eqnarray}
\ket{1}_S &=& \frac{1}{\sqrt{N}} \sum^N_{k=1} \ket{ \downarrow_1 \downarrow_2 \cdots \downarrow_{k-1} \uparrow_k \downarrow_{k+1} \cdots \downarrow_N} \\
\ket{2}_S &=& \frac{2}{\sqrt{N(N-1)}} \sum^N_{k=1} \sum^N_{l=1, l \neq k} \ket{ \downarrow_1 \downarrow_2 \cdots \downarrow_{k-1} \uparrow_k \downarrow_{k+1} \cdots \downarrow_{l-1} \uparrow{l} \downarrow{l+1} \cdots \downarrow_N}
\end{eqnarray}

These states are necessary since we are not able to determine specifically which atom is in the ground state or the excited state \cite{CAVQEDBIG}.  Applying the $\hat{J}_{\pm}$ to the symmetric states gives the following
 
\begin{eqnarray}
\hat{J}_+ \ket{0}_S &=& \ket{1}_S \\
\hat{J}_- \ket{1}_S &=& \ket{0}_S 
\end{eqnarray}

And the same can be done with $\ket{2}_S$

The Hamiltonian we will deal with from now on is
\begin{equation} \label{eq: WF Ham}
\hat{H} =   i \hbar g (  \hat{a}^{\dagger} \hat{J}_- - \hat{a} \hat{J}_+ ) + i \hbar Y ( \hat{a}^{\dagger} -  \hat{a})- i\hbar (\frac{\gamma}{2} \hat{J}_+ \hat{J}_- + \kappa \hat{a}^{\dagger} \hat{a})
\end{equation}
If we were to solve the Schroedinger equation from this Hamiltonian there would be an infinite number of states.  In order to truncate the basis we apply the weak field limit where there is less than two excitations (the number of photons plus the number of excited atoms is equal to the number of excitations.)
  
\begin{eqnarray}
\dot{C}_{00} &=& -Y C_{10} \\
\dot{C}_{10} &=& Y (C_{00} - C_{20})+ g\sqrt{N} C_{11} - \kappa C_{10} \\
\dot{C}_{01} &=& Y C_{11} - g\sqrt{N}  C_{00} - \frac{\gamma}{2} C_{01} \\
\dot{C}_{20} &=& \sqrt{2} Y C_{10} +g \sqrt{2} \sqrt{N} C_{11} - 2 \kappa C_{20} \\
\dot{C}_{11} &=&  Y C_{01} - g \sqrt{2} \sqrt{N} C_{01} + g \sqrt{2} \sqrt{N-1} C_{02} - (\kappa + \frac{\gamma}{2}) C_{20} \\
\dot{C}_{02} &=& - g \sqrt{2} \sqrt{N-1}C_{11} - \gamma C_{02}
\end{eqnarray}
  
Here $C_{nm}$ are the amplitudes of the $\ket{n m}$ state where $n$ is the number of photons and $m$ is the number of atoms in the excited state.  In the weak field limit, we have $C_{nm}\approx Y^{(n+m)/2}$. The excitation of the system is then referred to as the sum $n + m$.

This is, however, still not correct since we are missing higher order terms, with the truncated basis, that should appear with $\dot{C}_{20}$, $\dot{C}_{11}$, and $\dot{C}_{02}$.  In order to fix this problem we go back to the initial assumption of a weak field limit.    Therefore some terms multiplied by $Y^s$, where $s$ is some integer dependent on the state, do not affect the amplitude rates.  Then we can scale the amplitudes to be on order 1, $Y$, or $Y^2$ grouped by the excitation \cite{OptComm}.  We assume the ground state $C_{00} \sim 1$ therefore we drop the $C_{10}$ term in $\dot{C}_{00}$ since its at least of order $Y$.  From here it is clear to see that the first excited states ($C_{10}$ and  $C_{01}$) are order $Y$ and then we can drop any higher order terms since $Y$ is small) \cite{OptComm}.  Similarly the second excited states ($C_{20}$, $C_{11}$ and $C_{02}$) are of order $Y^2$ and so higher order $Y$ terms can be dropped.  The final states are 
   \label{eq: weak field amps}
\begin{eqnarray}
\dot{C}_{00} &=& 0\\
\dot{C}_{10} &=& Y + g\sqrt{N} C_{11} - \kappa C_{10} \\
\dot{C}_{01} &=&  - g\sqrt{N}  C_{00} - \frac{\gamma}{2} C_{01} \\
\dot{C}_{20} &=& \sqrt{2} Y C_{10} +g \sqrt{2} \sqrt{N} C_{11} - 2 \kappa C_{20} \\
\dot{C}_{11} &=&  Y C_{01} - g \sqrt{2} \sqrt{N} C_{01} + g \sqrt{2} \sqrt{N-1} C_{02} - (\kappa + \frac{\gamma}{2}) C_{20} \\
\dot{C}_{02} &=& - g \sqrt{2} \sqrt{N-1}C_{11} - \gamma C_{02}
\end{eqnarray}

These are the final amplitudes for the weak field trajectory model \cite{OptComm}.  We must remember, though, that in this weak field limit, where the amplitudes are scaled to 1, $Y$, $Y^2$,..., that all other quantities calculated from them must also be scaled \cite{OptComm}.  For example, the jump probability discussed earlier.

To make our model more efficient we will need to make some approximations.  First, we will describe adiabatic elimination to further simplify our Hamiltonian.  Next, we introduce photon counting statistics which will be our main tool in investigating the system.  

\section{Adiabatic Elimination} \label{ss: AE}
Adiabatic elimination is a method to determine the behavior of the system in different time intervals by invoking limits to the parameters involved.   We start here from the Hamiltonian used in the previous section, equation (\ref{eq: WF Ham}).  The limit we will use in this paper is the bad cavity limit meaning $\kappa$ dominates all other rates.  

We have
   \label{eq: ada rates}
\begin{eqnarray}
\dot{\hat{a}} &=& \frac{i}{\hbar} [ H,\hat{a}] = - g\hat{J}_- - Y + \kappa \hat{a} \\
\dot{\hat{a}}^{\dagger} &=& \frac{i}{\hbar} [ H,\hat{a}^{\dagger}] =  g\hat{J}_- + Y - \kappa \hat{a}^{\dagger}
\end{eqnarray}

The equations for $\dot{\hat{\sigma}}_-$ and $\dot{\hat{\sigma}}_+$ can be solved similarly, however, that is not the simplification we wish to study here \cite{Larson}.  Taking the limit where the rates of $\dot{\hat{a}} \rightarrow 0$ and $\dot{\hat{a}}^{\dagger} \rightarrow 0$ we can get the equations for $\hat{a}$ and $\hat{a}^{\dagger}$. 

\begin{eqnarray}
\hat{a} &=& \frac{g}{\kappa} \hat{J}_- + \frac{Y}{\kappa} \\
\hat{a}^{\dagger} &=& \frac{g}{\kappa} \hat{J}_+ + \frac{Y}{\kappa}
\end{eqnarray}

The Hamiltonian we derived previously comes from the density operator \cite{OptComm}.  As explained earlier the system is interacting with an external reservoir, therefore we cannot simply replace terms in equation (\ref{eq: WF Ham}) with our new expressions for $\hat{a}$ and $\hat{a}^{\dagger}$.  In order to derive the correct Hamiltonian we must start back with the density operator 
\begin{equation}
\dot{\rho} = -\frac{i}{\hbar} \left[ H, \rho \right] + \frac{\gamma}{2} \left[2 \hat{J}_- \rho \hat{J}_+ - \hat{J}_+ \hat{J}_-  \rho - \rho \hat{J}_+ \hat{J}_- \right] + \kappa \left[ 2 \hat{a} \rho \hat{a}^{\dagger} - \hat{a}^{\dagger} \hat{a}  \rho -  \rho \hat{a}^{\dagger} \hat{a} \right]
\end{equation}
Now we can substitute the new $\hat{a}$ and $\hat{a}^{\dagger}$ into the master equation.  Instead of substituting in the terms all at once, we will go through each term that is changed starting first with the interaction term in the Hamiltonian.  

The first part comes from the interaction term
\begin{eqnarray}
i \hbar g \left(\hat{a}^{\dagger} \hat{J}_- - \hat{a} \hat{J}_+ \right) &=& i \hbar g \left[ \left( \frac{g}{\kappa} \hat{J}_+ + \frac{Y}{\kappa} \right) \hat{J}_- - \left(\frac{g}{\kappa} \hat{J}_- + \frac{Y}{\kappa} \right) \hat{J}_+\right] \nonumber \\
		&=& i \hbar g \frac{Y}{\kappa} (\hat{J}_- - \hat{J}_+) + i \hbar \frac{g^2}{\kappa} (\hat{J}_+ \hat{J}_- - \hat{J}_- \hat{J}_+)
\end{eqnarray}
In this part the second term is equivalent to $\hat{J}_z$ and can be ignored since it only changes the zero point energy.  

The next term changed is the weak driving laser
\begin{equation}
i \hbar Y (\hat{a}^{\dagger} - \hat{a}) = i \hbar g \frac{Y}{\kappa} (\hat{J}_+ - \hat{J}_-) + i \hbar\frac{Y^3}{\kappa^2}
\end{equation}
Again, since the final term is higher, it can be ignored.  The first term is equal but opposite to the first term coming from the interaction term and so, they eliminate each other.  This may be disconcerting at first but we must remember that there are more parts from the $\kappa$ portion of equation for $\dot{\rho}$.  

The $\kappa$ dissipation term has three parts from the foiling of the equations for $\hat{a}$ and $\hat{a}^{\dagger}$
\begin{eqnarray}
\kappa \left[ 2 \hat{a} \rho \hat{a}^{\dagger} -  \hat{a}^{\dagger} \hat{a} \rho - \rho \hat{a}^{\dagger} \hat{a} \right] &=& 
 \frac{g^2}{\kappa} \left(2 \hat{J}_- \rho \hat{J}_+ - \hat{J}_+ \hat{J}_-  \rho - \rho \hat{J}_+ \hat{J}_-\right)\nonumber \\
&& + \frac{i}{\hbar} \left[i \hbar \frac{g Y}{\kappa} (\hat{J}_- - \hat{J}_+), \rho \right] 
\end{eqnarray}
The first part comes from the $g \hat{J}/\kappa$ while the second comes from the cross term between the $\hat{J}$ and the constant term.  There was a third term, which was constant, which we again eliminate by shifting the zero point energy.  The first term has the same operation on $\rho$ as the $\gamma$ term and so we can combine the two to give a new decay term in our Hamiltonian or letting $\gamma \rightarrow \frac{\gamma}{2}( 1 + \frac{2 g^2}{\kappa \gamma})$.  We move the second part to the evolution Hamiltonian to create the bad cavity Hamiltonian
\begin{equation} \label{eq: BC Ham}
\hat{H}_{BC} = i \hbar \frac{g Y}{\kappa} \left(\hat{J}_- - \hat{J}_+ \right) - \frac{\gamma}{2} \left( 1 + 2 \frac{g^2}{\kappa \gamma} \right) \hat{J}_+ \hat{J}_- 
\end{equation}
This Hamiltonian is much simpler to work with since we eliminated the field variables.  By using this method, we can solve analytically for some situations.  The probability of a jump from transmission is also edited because of the substitution for $\hat{a}$ and $\hat{a}^{\dagger}$ as well as the collapse operator.
\begin{equation} \label{eq: BC jump prob}
P_{\kappa} = 2 \kappa \left( \frac{Y^2}{\kappa^2} + \frac{g Y}{\kappa^2} (\braket{\hat{J}_+} + \braket{\hat{J}_-})+ \frac{g^2}{\kappa^2} \braket{\hat{J}_+ \hat{J}_-}\right) \Delta t
\end{equation}
Each term corresponds to a different type of emission \cite{OptComm}.  The first is from photons passing through the cavity without being absorbed by an atom.  The second term is from coherent emission by the atoms into the cavity mode.  The third term is from cavity enhanced spontaneous emission of the atoms into the cavity mode.

\section{Two-component BEC model} \label{ss: 2comp BEC}
We can now combine the weak field trajectory terms, equation (\ref{eq: 2comp Ham}), with the Bose-Hubbard Hamiltonian, to get a new Hamiltonian describing the evolution of a two level boson atom in an optical lattice.  The Hamiltonian for the system is
\begin{eqnarray} \label{eq: 2comp BEC Ham}
H &=& - \hbar J \sum_{i,j} \hat{b}^{\dagger}_i \hat{b}_j + \frac{U}{2} \sum_i \hat{n}_i \left(\hat{n}_i -1 \right) +  i \hbar g \left(  \hat{a}^{\dagger} \hat{J}_- - \hat{a} \hat{J}_+\right) + i \hbar Y \left( \hat{a}^{\dagger} -  \hat{a}\right) \nonumber \\
&& - i\hbar \left(\frac{\gamma}{2} \hat{J}_+ \hat{J}_- + \kappa \hat{a}^{\dagger} \hat{a}\right)
\end{eqnarray}
The Hamiltonian allows boson atoms to travel between lattice sites and also interact with a weak driving laser $Y$.  

The number of states for this system is the product of the number of states from the Bose-Hubbard model and the the number of states from the trajectory  model.  The number of states from the Bose-Hubbard Hamiltonian is proportional to the number of sites, $L$ and the number of particles $N$ \cite{Arnett}
\begin{equation}
S_{BH} = \frac{(N + L - 1)!}{N!(L - 1)!}
\end{equation}
The number of states for the trajectory Hamiltonian is proportional to the number of particles and the number of excitations.  In the weak field limit there are six possible states for $N$ atoms and five for a single atom.   Then the total number of states is 
\begin{equation}
S_{tot}=6 \times \frac{(N + L - 1)!}{N!(L - 1)!}
\end{equation}
For example, when there are two sites and one particles there are two states from the atom configuration ($\ket{10}$ and $\ket{01}$ where a 1 denotes the one atom in a site) multiplied by the different trajectory states which for the weak filed limit (two excitations) is five.  This gives ten possible states. 

\begin{center}
\begin{figure} \label{fig: 2 site}
\includegraphics{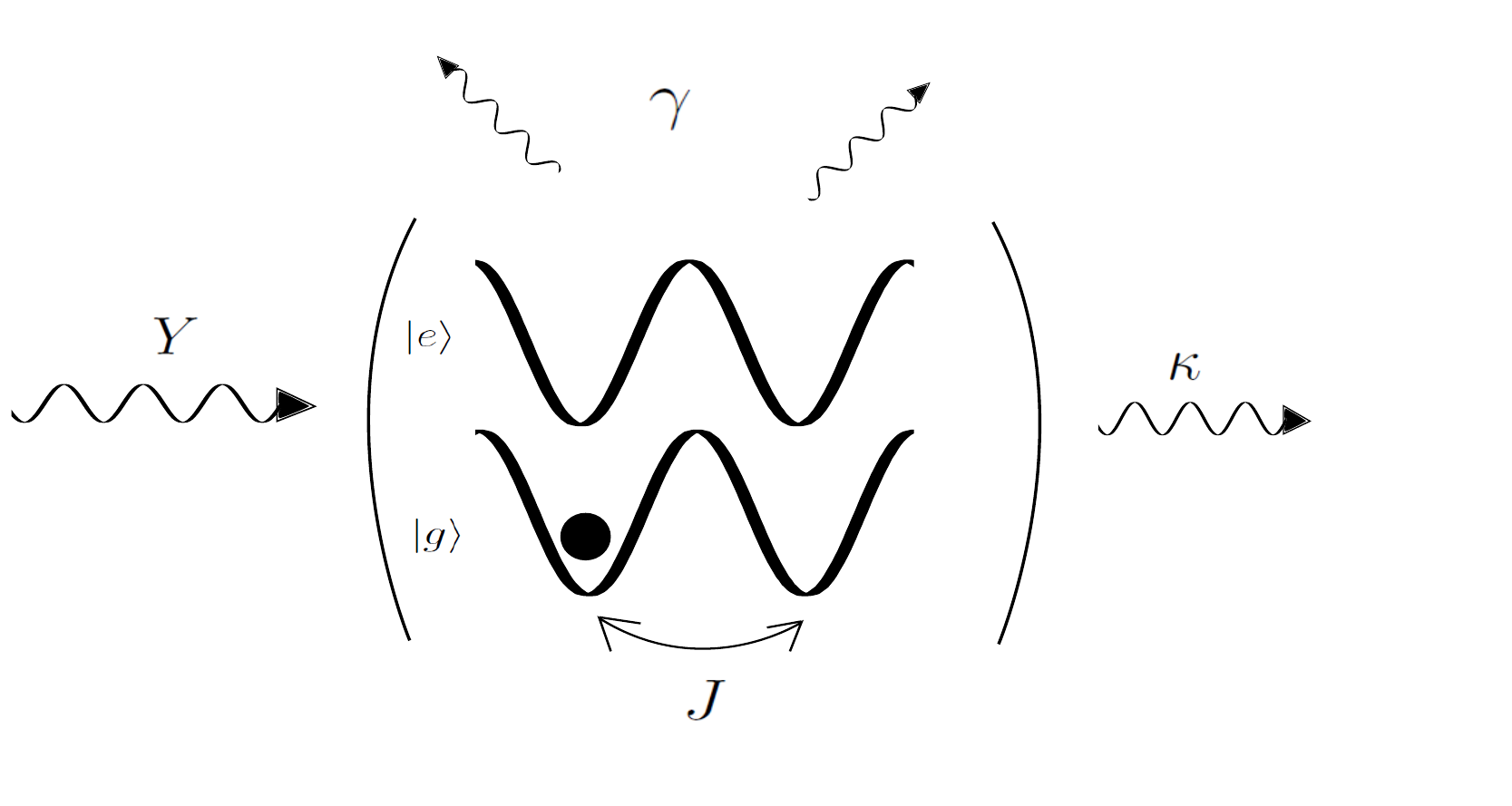}
\caption[Two site one atom setup]{This is the two lattice site setup with the two-component model where $Y$ is the driving laser, $\gamma$ is spontaneous emission, $\kappa$ the cavity loss and $J$ the tunneling coefficient}
\end{figure}
\end{center}

There is also the bad cavity Hamiltonian found by adiabatic elimination of a $\hat{a}$ and $\hat{a}^{\dagger}$ of the trajectory model as shown in section (\ref{ss: AE}). 
\begin{equation} \label{eq: 2comp BEC BC Ham}
H_{BC} =- \hbar J \sum_{i,j} \hat{b}^{\dagger}_i \hat{b}_j + \frac{U}{2} \sum_i \hat{n}_i (\hat{n}_i -1) +  i \hbar \frac{g Y}{\kappa} \left(\hat{J}_- - \hat{J}_+ \right) - \frac{\gamma}{2} \left( 1 + 2 \frac{g^2}{\kappa \gamma}\right)\hat{J}_+ \hat{J}_- 
\end{equation}
In this case the number of states based on the configuration of atoms is the same as the previous Hamiltonian, however, the number of states from the bad cavity Hamiltonian is only equal to the number of excited states since $\hat{a}$ and $\hat{a}^{\dagger}$ were eliminated.  
\begin{equation}
S_{tot}= (e_{max} + 1) \times \frac{(N + L - 1)!}{N!(L - 1)!}
\end{equation}
Where the "$+1$" represents the ground state.  For the weak field limit $e_{max} =2$ for $N$ atoms while for a single atom $e_{max}=1$.   Therefore, for the two sites one particle arrangement there are only four possible states.  The number of states from the atom configuration are multiplied by two now.  For the following sections we will restrict our analysis to this system of a single atom allowed to jump between two lattice sites as shown in figure (\ref{fig: 2 site}).

Now we analyze the single atom two-cavity system with the bad cavity Hamiltonian.  We use the bad cavity Hamiltonian to solve Schroedinger's equation to get the rate of change of the amplitudes.  The $U$ term is dropped since that term only appears when there are more than one atoms.  This also means that we are not considering an actual BEC since there are not multiple atoms to interact. 

We use the notation $C_{nm}^i$ where the subscript shows the number of atoms in site one ($n$) and the number of atoms in site two ($m$) and the superscript denotes the internal energy $i=g,e$.  Solving Schroedinger's equation gives

\begin{eqnarray}
\dot{C}_{10}^g &=& i J C_{01}^g + \frac{g Y}{\kappa} C_{10}^e  \\
\dot{C}_{01}^g &=& i J C_{10}^g + \frac{g Y}{\kappa} C_{01}^e \\
\dot{C}_{10}^e &=& i J C_{01}^e - \frac{g Y}{\kappa} C_{10}^g - \frac{\gamma}{2} (1 + 2 C) C_{10}^e \\
\dot{C}_{01}^e &=& i J C_{10}^e - \frac{g Y}{\kappa} C_{01}^g - \frac{\gamma}{2} (1 + 2 C) C_{01}^e 
\end{eqnarray}

Where $C = g^2/\kappa \gamma$.  From here we can solve the equations numerically or take a weak field limit and solve them analytically.  The numerical solution requires a Runge-Kutta method instead of an Euler method.

The amplitude rates are scaled on $Y$ similar to in the simple weak field trajectory model shown in section (\ref{ss: 2comp}.)  The ground states are of order 1 and the excited states are of order $Y$.  This results in dropping the $Y$ terms in the ground state rates

\begin{eqnarray}
\dot{C}_{10}^g &=& i J C_{01}^g   \\
\dot{C}_{01}^g &=& i J C_{10}^g \\
\dot{C}_{10}^e &=& i J C_{01}^e - \frac{g Y}{\kappa} C_{10}^g - \frac{\gamma}{2} (1 + 2 C) C_{10}^e \\
\dot{C}_{01}^e &=& i J C_{10}^e - \frac{g Y}{\kappa} C_{01}^g - \frac{\gamma}{2} (1 + 2 C) C_{01}^e 
 \label{eq: weak field Amps}
\end{eqnarray}

These equations are coupled so to separate them we use the symmetry states labeled $D^{g,e}_\pm$ which relate to $C^{g,e}$ by

\begin{eqnarray}
D_+^g &=& \frac{1}{\sqrt{2}} ( C_{10}^g + C_{01}^g) \\
D_-^g &=& \frac{1}{\sqrt{2}} ( C_{10}^g - C_{01}^g) \\
D_+^e &=& \frac{1}{\sqrt{2}} ( C_{10}^e + C_{01}^e) \\
D_-^e &=& \frac{1}{\sqrt{2}} ( C_{10}^e - C_{01}^e) 
\end{eqnarray}

Substituting the $D$ equations into equation (\ref{eq: weak field Amps}) gives a series of uncoupled differential equations which can be solved analytically.  Undoing the substations for $D$ gives the amplitudes for the weak field bad cavity limit as

\begin{eqnarray}
C_{10}^g(t) &=& \frac{1}{\sqrt{2}} (D_+^g(0) e^{i J t} + D_-^g(0) e^{-i J t}) \\
C_{01}^g(t) &=& \frac{1}{\sqrt{2}} (D_+^g(0) e^{i J t} - D_-^g(0) e^{-i J t}) \\
C_{10}^e(t) &=& \left( \frac{Y D_+^g(0)}{\Gamma} (e^{-\Gamma t} -1) + D_+^e(0) e^{-\Gamma t}\right) \frac{e^{i J t}}{\sqrt{2}}  \nonumber\\ 
&& + \left( \frac{Y D_-^g(0)}{\Gamma} (e^{-\Gamma t}-1) + D_-^e(0) e^{-\Gamma t} \right) \frac{e^{-i J t}}{\sqrt{2}}  \\
C_{01}^e(t) &=&\left( \frac{Y D_+^g(0)}{\Gamma} (e^{-\Gamma t}-1) + D_+^e(0) e^{\Gamma t} \right)\frac{ e^{i J t}}{\sqrt{2}} \nonumber \\
 &&-  \left( \frac{Y D_-^g(0)}{\Gamma} (e^{-\Gamma t}-1) + D_-^e(0) e^{-\Gamma t} \right) \frac{e^{-i J t}}{\sqrt{2}}
\end{eqnarray}

We let $\gamma(1 + 2C)/ = \Gamma$.  These amplitudes only apply when $Y$ is small due to the weak field limit.  Therefore, we have two separate models to compare, one numeric and one analytic.  To simplify the solution we can restrict the analysis by applying the initial condition that the particle must start in the ground state of the first well.  Then, $D_+^g(0) = D_-^g(0) = 1/\sqrt{2}$ and $D_+^e(0) = D_-^e(0) = 0$.  Then the amplitudes are

\begin{eqnarray}
C_{10}^g(t) &=& \mbox{cos}(J t) \\
C_{01}^g(t) &=& \mbox{sin}(J t) \\
C_{10}^e(t) &=& \frac{Y}{\Gamma} \left( e^{-  \Gamma t} - 1 \right) \mbox{cos}(Jt) \\
C_{01}^e(t) &=& \frac{Y}{\Gamma} \left( e^{-  \Gamma t} - 1 \right) \mbox{sin}(J t)
 \label{eq: C amps}
\end{eqnarray}

Starting the particle in the other well will only shift the phase of each by $\pi/2$.  This will be the starting condition used throughout.

We start the system in the ground state with the atom in the first well.  The program then does a series of stpdf at a set time interval.  At each step the probability of a jump is calculated,  for spontaneous emission or cavity loss.  For this case (2 sites 1 atom,) there are three jumps.  Two for spontaneous emission (one for each sites) and one for cavity loss.  We can also look at a jump from either site, which is the sum of the jumps from each site.
  
\begin{eqnarray}
P_{\gamma_1} &=& \gamma | C_{10}^e |^2 \Delta t \\
P_{\gamma_2} &=& \gamma | C_{01}^e |^2 \Delta t \\
P_{\gamma} &=& P_{\gamma_1} + P_{\gamma_1} =  \gamma \left( | C_{10}^e |^2 + | C_{10}^e |^2 \right) \Delta t \\
P_{\kappa}   &=& 2 [ \frac{Y^2}{\kappa} + \frac{Y g}{\kappa} ({C_{10}^e}^*C_{10}^g + {C_{01}^e}^*C_{01}^g + {C_{10}^g}^*C_{10}^e + {C_{01}^g}^*C_{01}^e) \nonumber \\
 &&+\frac{g^2}{\kappa} ( |C_{10}^e |^2 + |C_{01}^e|^2)] \Delta t
\end{eqnarray}

The program then generates a random number and if it falls into a range determined by the probabilities calculated above a jump occurs.  The time of the jump is stored and the collapse operators collapse the system

\begin{eqnarray}
\hat{\sigma}_-^1 \ket{\psi} &=& C_{10}^e \ket{10g} \\
\hat{\sigma}_-^2 \ket{\psi} &=& C_{01}^e \ket{01g} \\
\hat{a} \ket{\psi} &=& \left(\frac{Y}{\kappa} C_{10}^g + \frac{g}{\kappa} C_{10}^e\right) \ket{10g} +  \left(\frac{Y}{\kappa} C_{01}^g + \frac{g}{\kappa} C_{01}^e\right) \ket{01g} \nonumber \\
&&+\frac{Y}{\kappa}C^e_{10} \ket{10e} + \frac{Y}{\kappa} C^e_{01}\ket{01e}
\end{eqnarray}

If however the random number does not fall into the jump range then the system will evolve.  The numeric and weak field analytic methods differ in this evolution.  For the numerical evolution, the program uses the Schroedinger equation solutions with a Runge-Kutta method.  The analytical model uses the solutions for the amplitudes.  With either case the final amplitudes are renormalized since the Hamiltonian's are non-Hermitian.  
Ultimately, the data taken will be the times of the jumps but to check the model we can  the expectation values $\braket{\hat{\sigma}_+^i \hat{\sigma}_-^i}$ and $\braket{\hat{a}^{\dagger} \hat{a}}$ which correspond to the probability of each jump.

\begin{eqnarray}
\braket{\hat{\sigma}_+^1 \hat{\sigma}_-^1} &=& |C_{10}^e|^2 \\
\braket{\hat{\sigma}_+ \hat{\sigma}_-} &=&  |C_{10}^e|^2 +  |C_{01}^e|^2 \\
\braket{\hat{a}^{\dagger} \hat{a}} &=&  \frac{Y^2}{\kappa} + \frac{Y g}{\kappa} ({C_{10}^e}^*C_{10}^g + {C_{01}^e}^*C_{01}^g + {C_{10}^g}^*C_{10}^e + {C_{01}^g}^*C_{01}^e) \nonumber \\
 &&+\frac{g^2}{\kappa} ( |C_{10}^e |^2 + |C_{01}^e|^2)
\end{eqnarray}

Which are plotted below with $J=1$, $g=\sqrt{10}$, $\kappa=10$, $\gamma=1$ so $C=1$.  Each type of jump is labeled to show how the collapse affects the expectation value of the function.
\begin{figure}[H]
\includegraphics{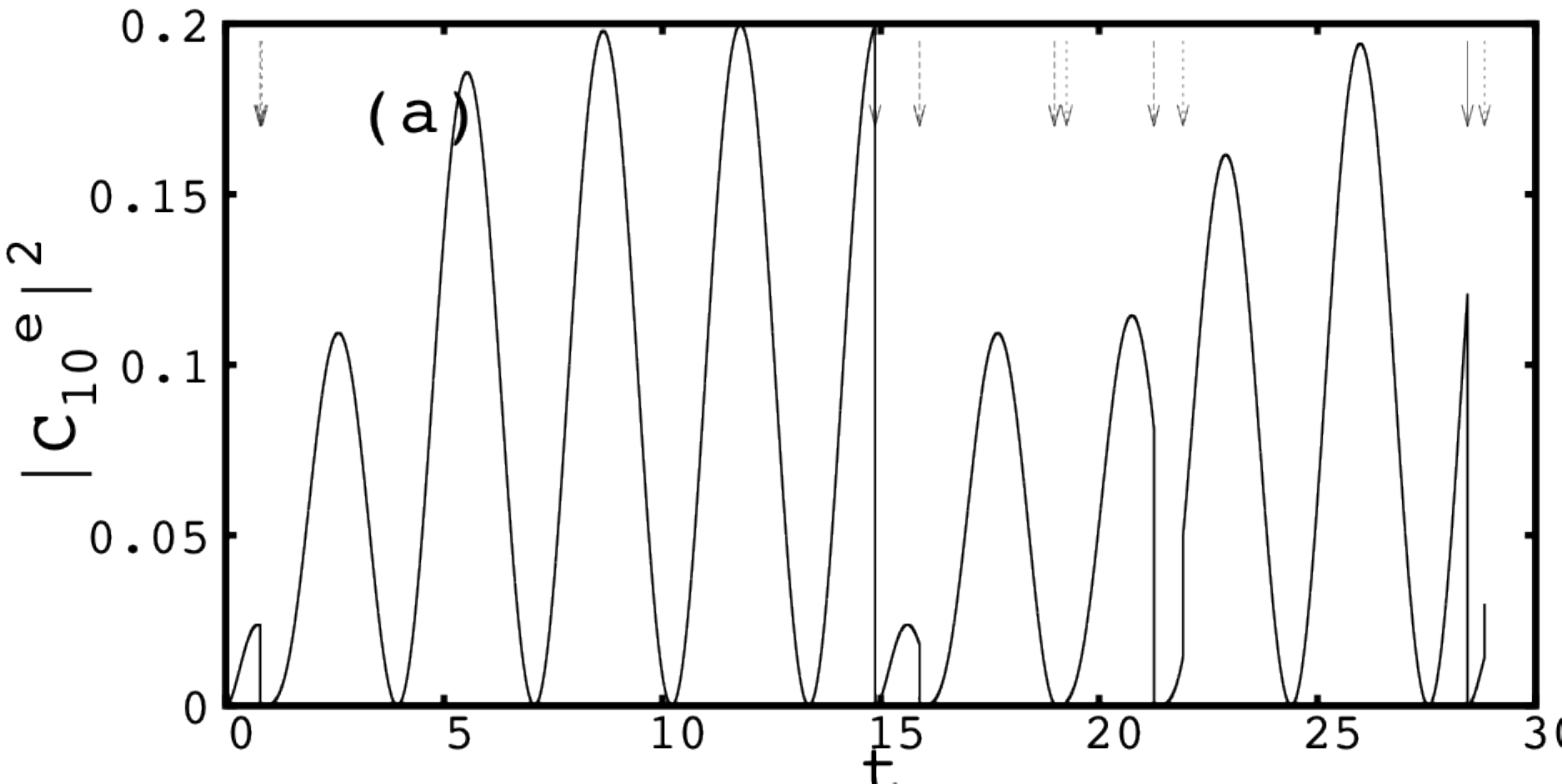}
\includegraphics{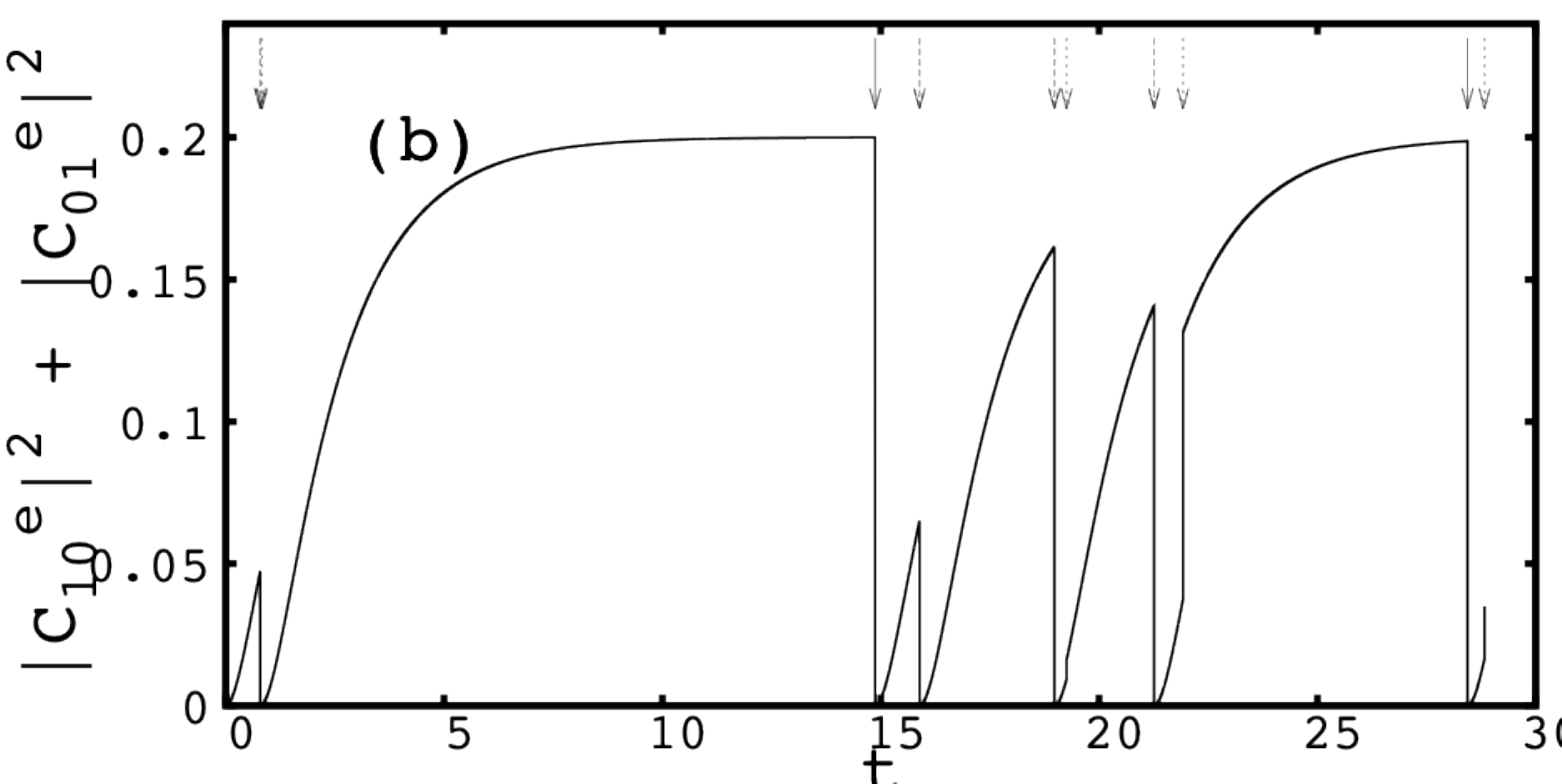}
\includegraphics{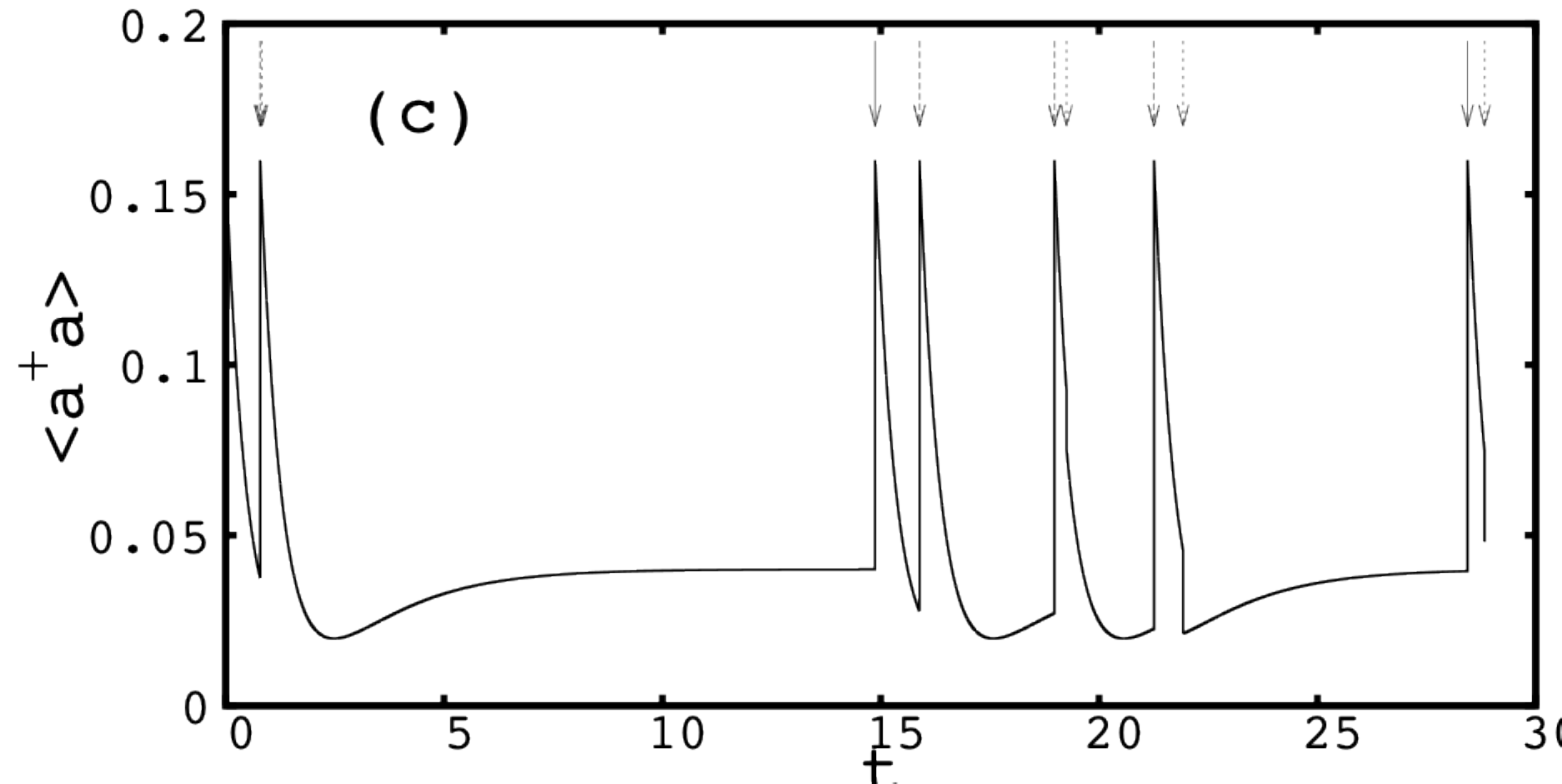}
\caption[Two stie one atom probability amplitudes with jumps]{A solid arrow shows a collapse in the first site, a dashed arrow a collapse in the second site and a finely dashed arrow a collapse from a transmitted photon.  (a) shows $\braket{\hat{\sigma}_+^1 \hat{\sigma}_-^1}$, (b) shows $\braket{\hat{\sigma}_- \hat{\sigma}_-}$ and (c) shows $\braket{\hat{a}^{\dagger} \hat{a}}$.}
\label{fig: collapses}
\end{figure}
The figure shows how each collapse operator affects the states differently.  For instance, a cavity transmission collapse does not bring the excited state to zero like a fluorescence collapse.  Since we do not know if the transmitted photon comes from the atom or the driving laser we do not know the state of the atom exactly.  Figure (\ref{fig: collapses}b) looks like the probability amplitude graphs seen in \cite{OptComm}.  The amplitude is not dependent on the site and therefore we cannot see any tunneling effects.  Figure (\ref{fig: collapses}a) looks the same but with oscillations overlaid.  This is the effect of adding tunneling.  However, we cannot measure the wave function to see the tunneling coefficient.  To gain information on the system we look at the correlation function $g^{(2)}(\tau)$ as explained in the next section.

\section{Results}
There are three lists of jump times output by each program, one list for each operator, spontaneous emission from each site and emission from the cavity.  For most trials the driving field, $Y$, was turned up in order to create many jumps.  This resulted in roughly $10^4$ jumps in $10^8$ iterations of the process described above. We vary the tunneling coefficient between trials to show differences in the results.  All other constants are kept at unity ($g=\sqrt{10}$, $\kappa=10$, $\gamma=1$ so $C=1$.)  Using these times we can determine the time between jumps, $\tau$ and plot the data in a histogram.  This relates to the counting statistic $g^{(2)}(\tau)$ except without normalization.  Still, from the histogram data it is possible to gain information about the tunneling coefficient $J$.  

There are eight possible types of histograms to create depending on what we count as the starting and stopping time for $\tau$.  The start and stops are coincide with a jump.  For this paper we focus on four of the possible eight; starting and stopping with a jump from the first site, starting with a jump from the first and stopping with a jump from the second, starting and stopping with a count from either site, and starting and stopping from a jump from the side of the cavity.

The times output by the models are the times of jumps to be used to calculated the time between jumps $\Delta t_j$.  We not only want to have  $\Delta t_j$ be a list of consecutive counts but also include counts from up to four jumps away to include any possible correlations.  We then let $\tau=\gamma \Delta t$ which gives $\tau$ dimensionless units.  The result is a list of $\tau$ values which can be binned to form a  histogram.

 In the numerical method, for counts measuring between two sites there are $\sim 2 \times 10^3$, measuring in a single site there are $\sim 10^4$, and measuring a count from either site or a transmission there are $\sim 3 \times 10^4$ values of $\tau$ in our range.  For the weak field analytical method, the counts are of order 10 less than the numerical method.  This is because we neglect the $Y$ term that drives the atoms to the excited state.  After this time the correlations are defined by classical probability and go to zero as $t \rightarrow \infty$.  We normalize the histogram counts to be independent of the number of jumps by dividing by the ratio of the total number of $\tau$'s to the number of bins.  This is not exaclty the correlation function, however, it is closely related so we will label it as such.

Here we present the primary data of the analysis shown above.  Each program was run for $2 \times 10^8$ iterations with $J=1,2,5$, $Y=0.4$ and $g=\sqrt{10}$, $\kappa=10$, $\gamma=1$ so $C=1$.  The data presented below is created by the method described in the previous section where $\tau<10$ for each type of correlation.  The results of the analytical model from the weak field analytical solution appear dashed while the results from the numerical Runge-Kutta method appear as solid lines.  The graphs go in order of increasing $J$ and appear with site correlations and one page and non-site dependent, correlations on the next.  

\begin{center}
\begin{figure}[H]
\includegraphics{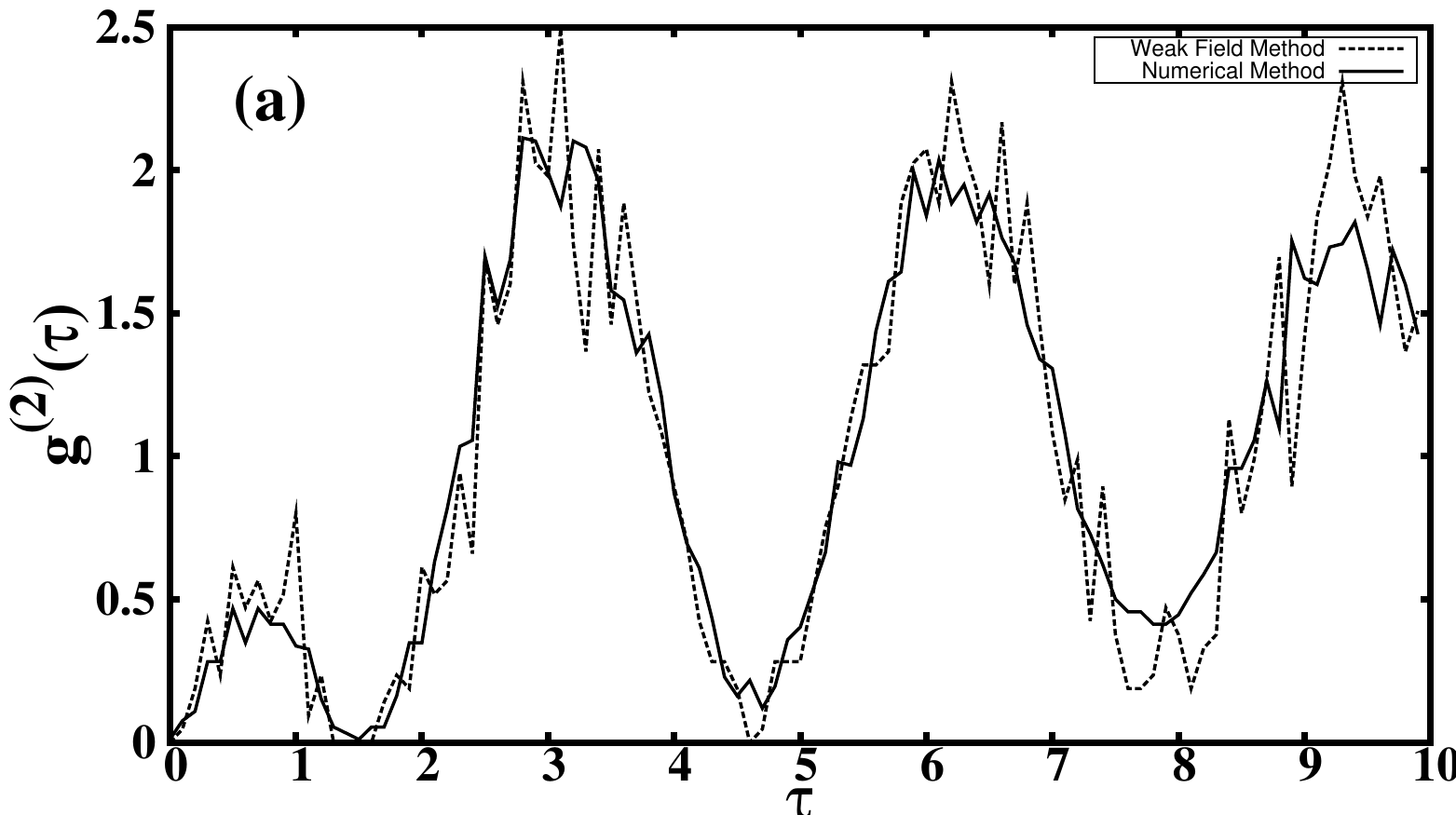} 
\\\\
\includegraphics{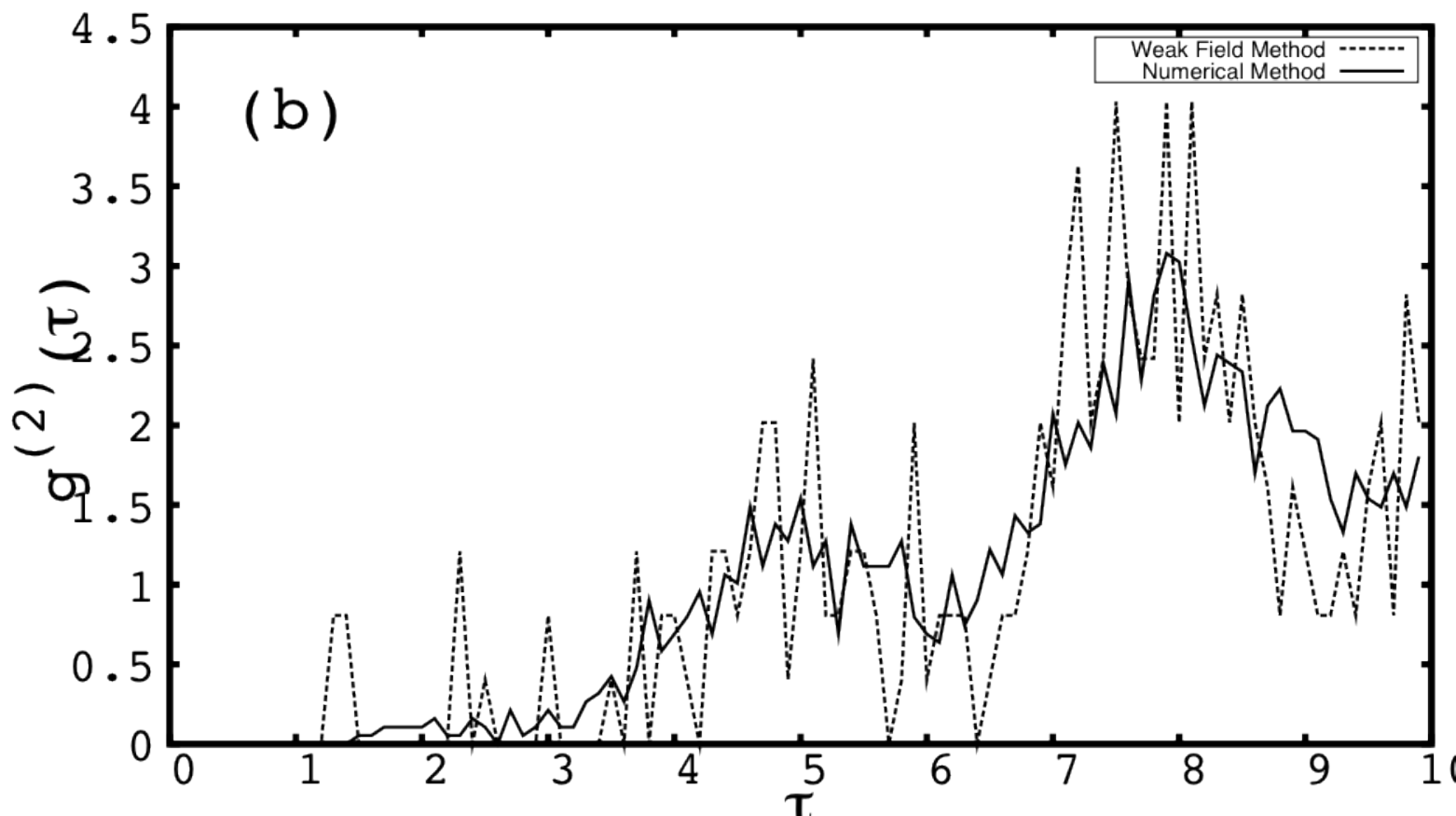}
\caption[Two-site numerical correlation functions for $J=1$]{Site correlation for $J=1$ (a) shows the correlation between photons fluoresced from a single site (b) shows the correlations between photons fluoresced from site 1 with those emitted from site 2}
\label{fig: T=1 sites}
\end{figure}
\end{center}

\begin{center}
\begin{figure}[H]
\includegraphics{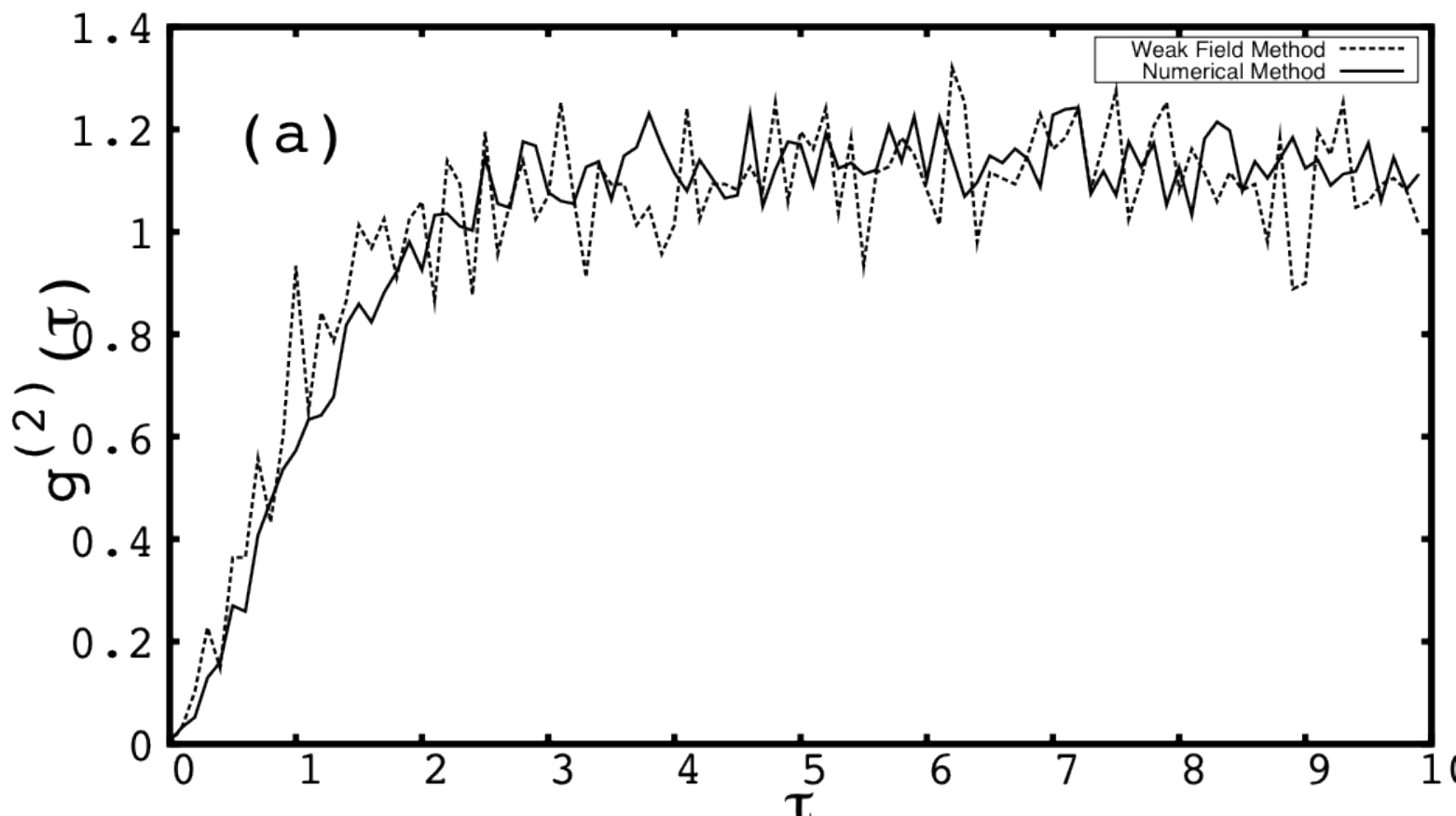}
\\\\
\includegraphics{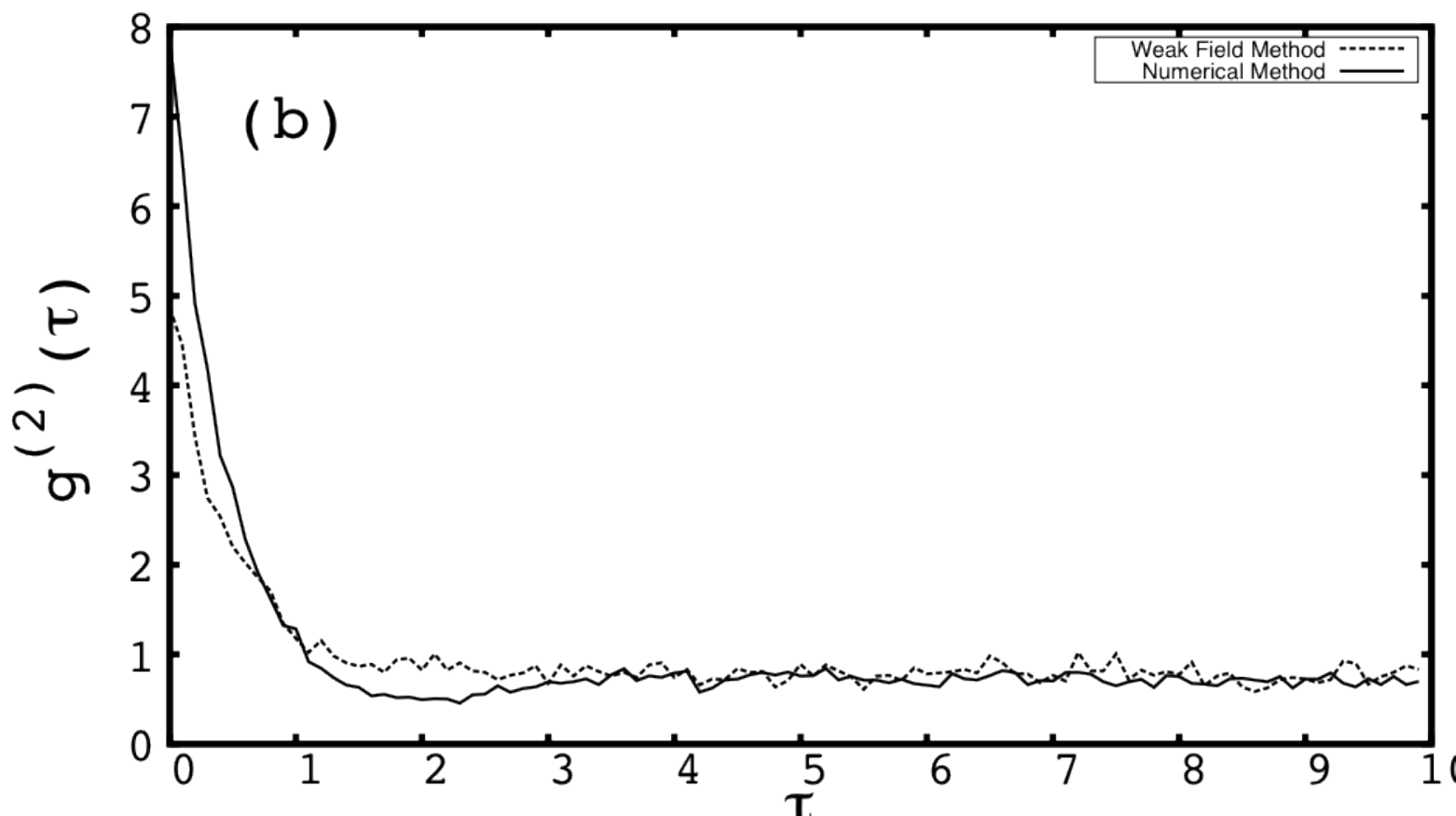}
\caption[Either site and transmission numerical correlation functions for $J=1$]{Site correlation for $J=1$ (a) shows the correlation between photons fluoresced regardless of site (b) shows the correlations between photons transmitted out the side of the cavity}
\label{fig: T=1 norm}
\end{figure}
\end{center}

\begin{center}
\begin{figure}[H]
\includegraphics{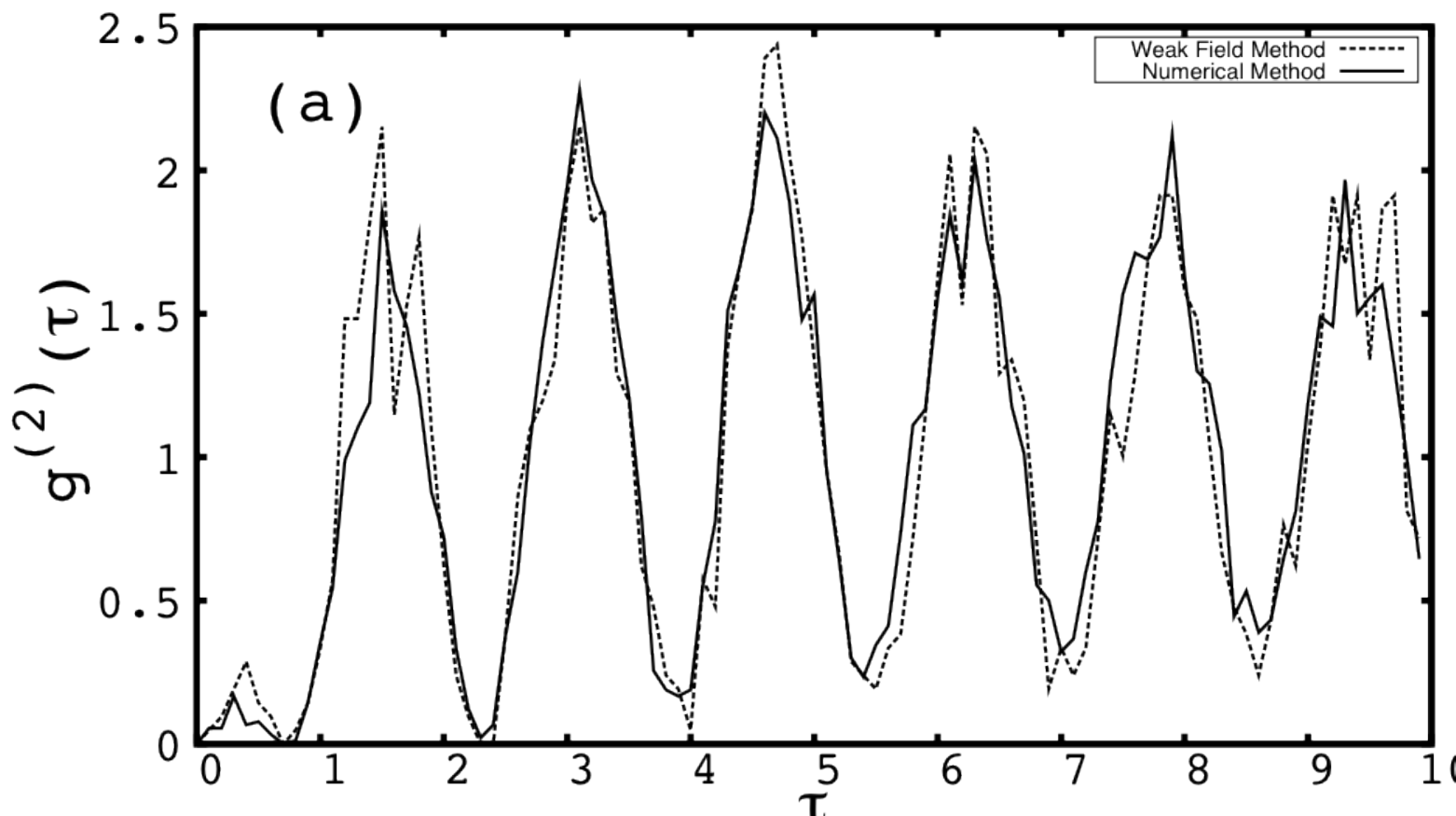} 
\\\\
\includegraphics{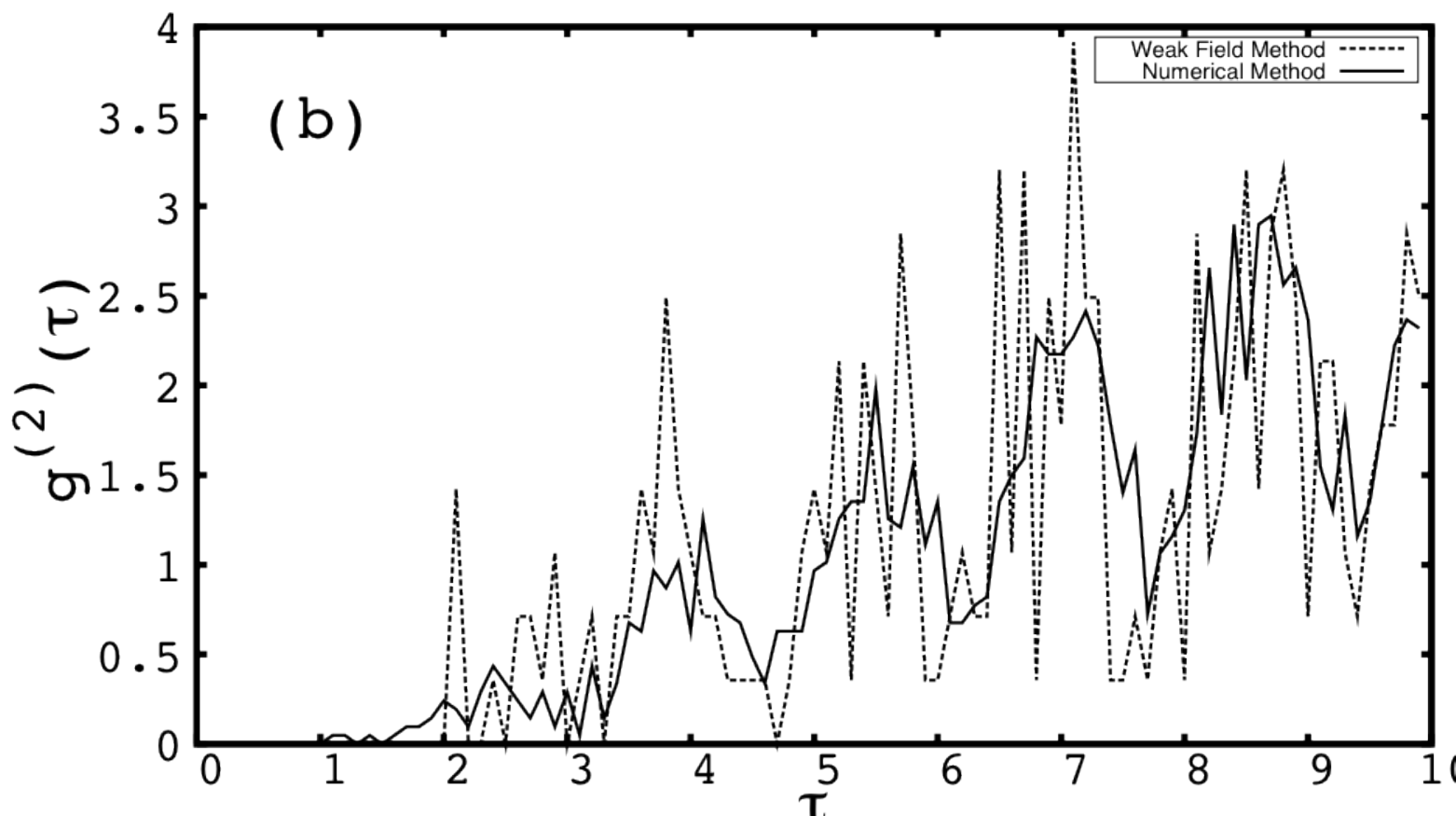}
\caption[Two-site numerical correlation functions for $J=2$]{Site correlation for $J=2$ (a) shows the correlation between photons fluoresced from a single site (b) shows the correlations between photons fluoresced from site 1 with those emitted from site 2}
\label{fig: T=2 sites}
\end{figure}
\end{center}

\begin{center}
\begin{figure}[H]
\includegraphics{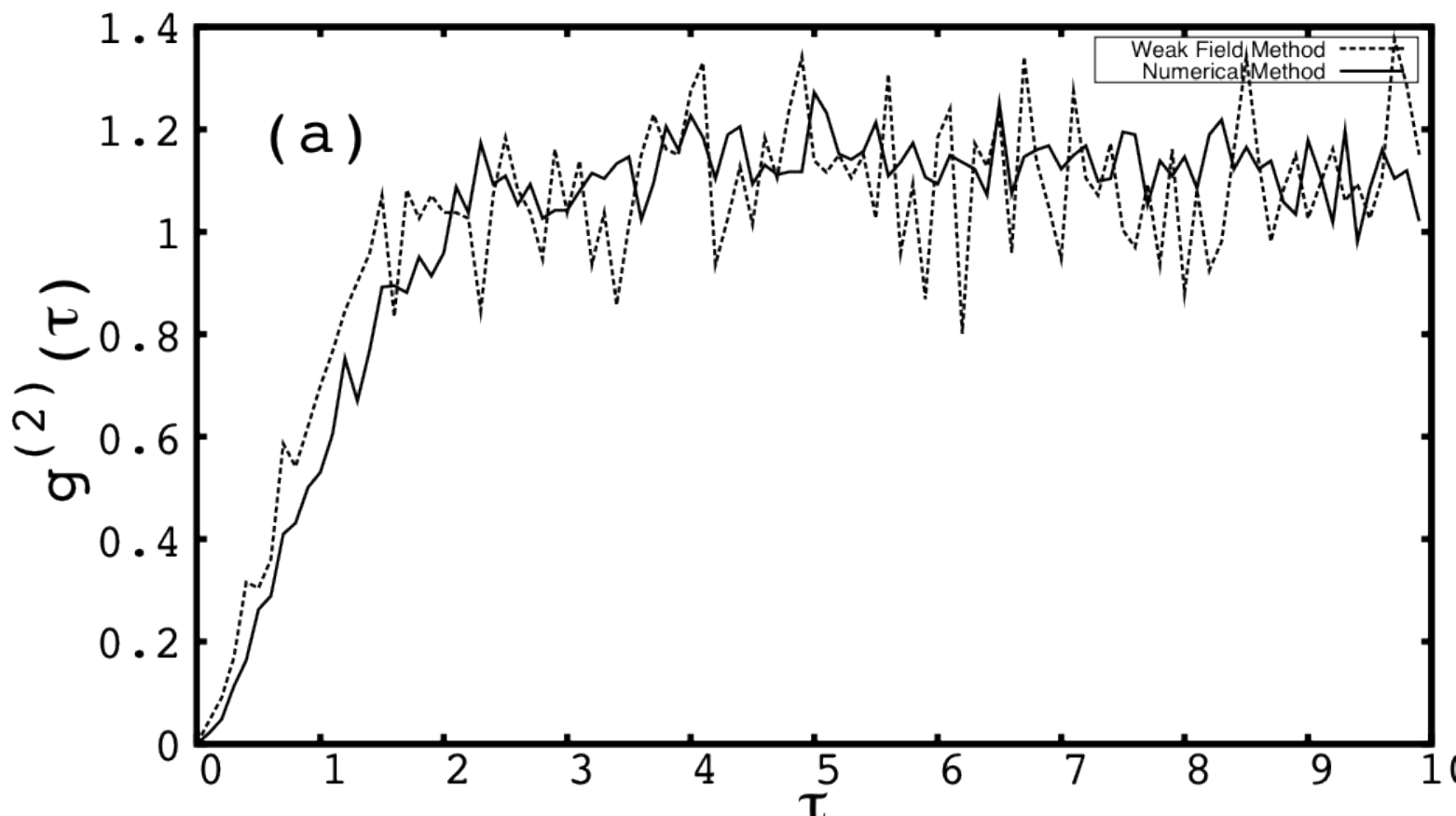}
\\\\
\includegraphics{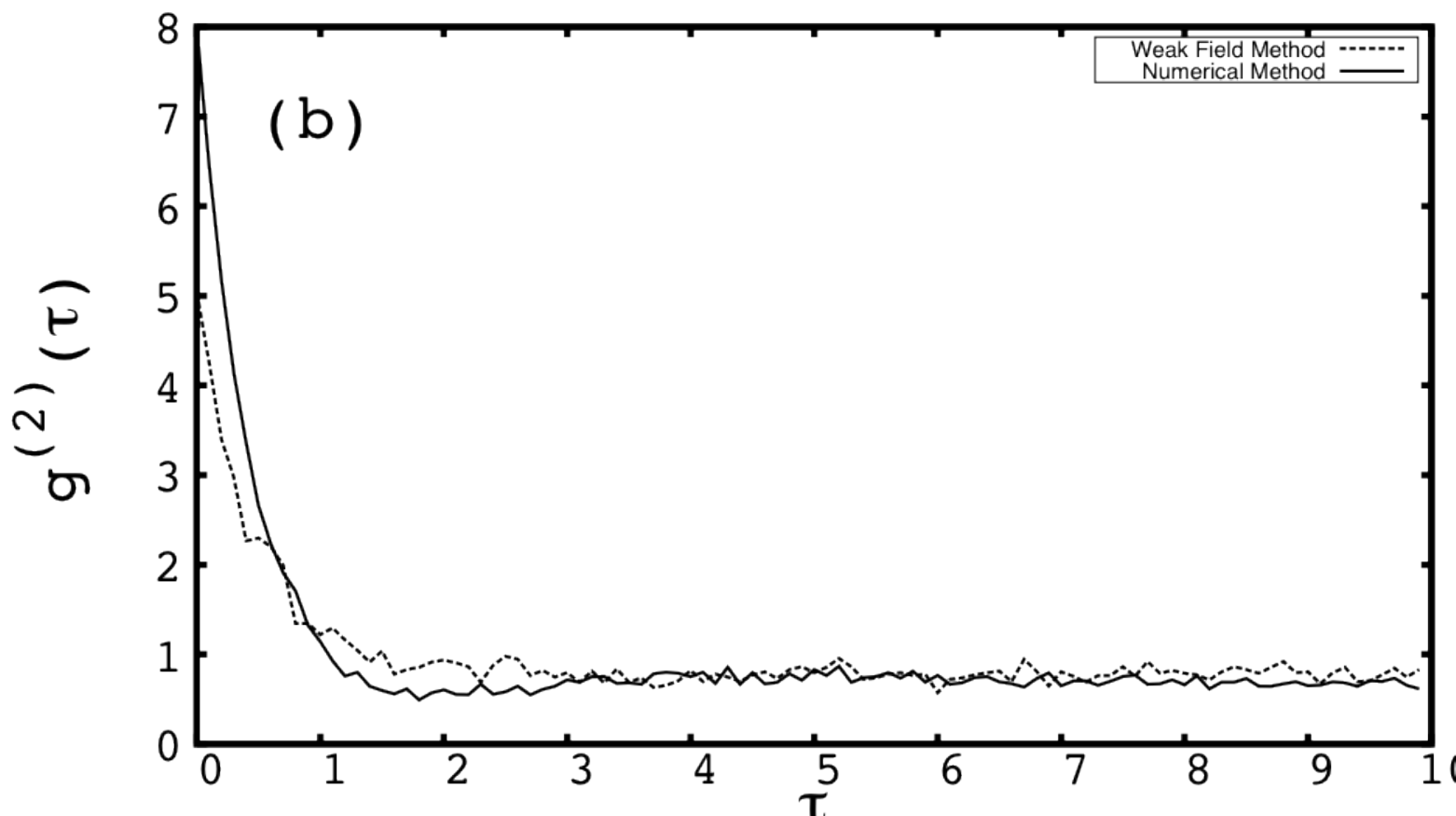}
\caption[Either site and transmission numerical correlation functions for $J=2$]{Site correlation for $J=2$ (a) shows the correlation between photons fluoresced regardless of site (b) shows the correlations between photons transmitted out the side of the cavity}
\label{fig: T=2 norm}
\end{figure}
\end{center}

\begin{center}
\begin{figure}[H]
\includegraphics{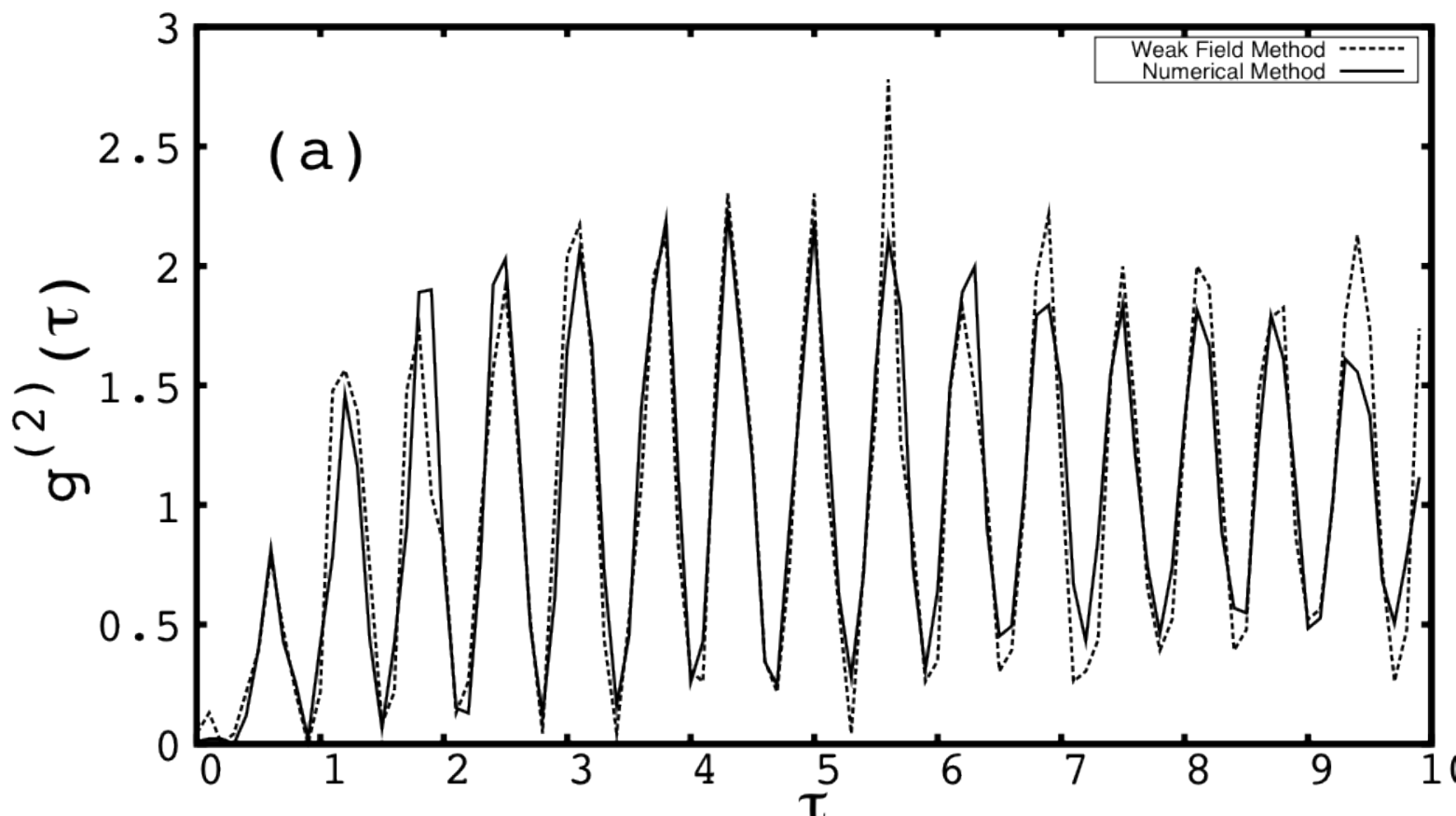} 
\\\\
\includegraphics{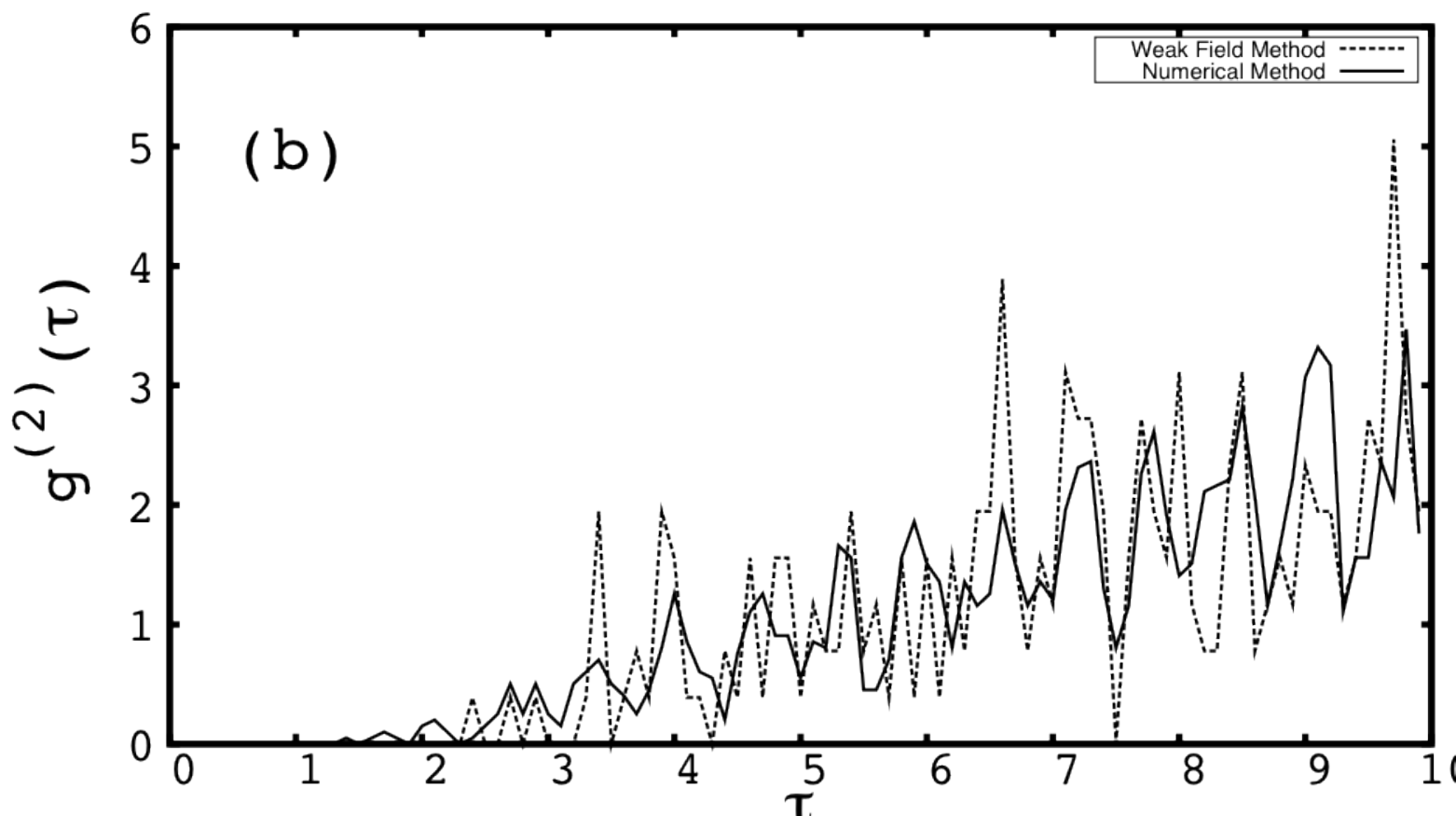}
\caption[Two-site numerical correlation functions for $J=5$]{Site correlation for $J=5$ (a) shows the correlation between photons fluoresced from a single site (b) shows the correlations between photons fluoresced from site 1 with those emitted from site 2}
\label{fig: T=5 sites}
\end{figure}
\end{center}

\begin{center}
\begin{figure}[H]
\includegraphics{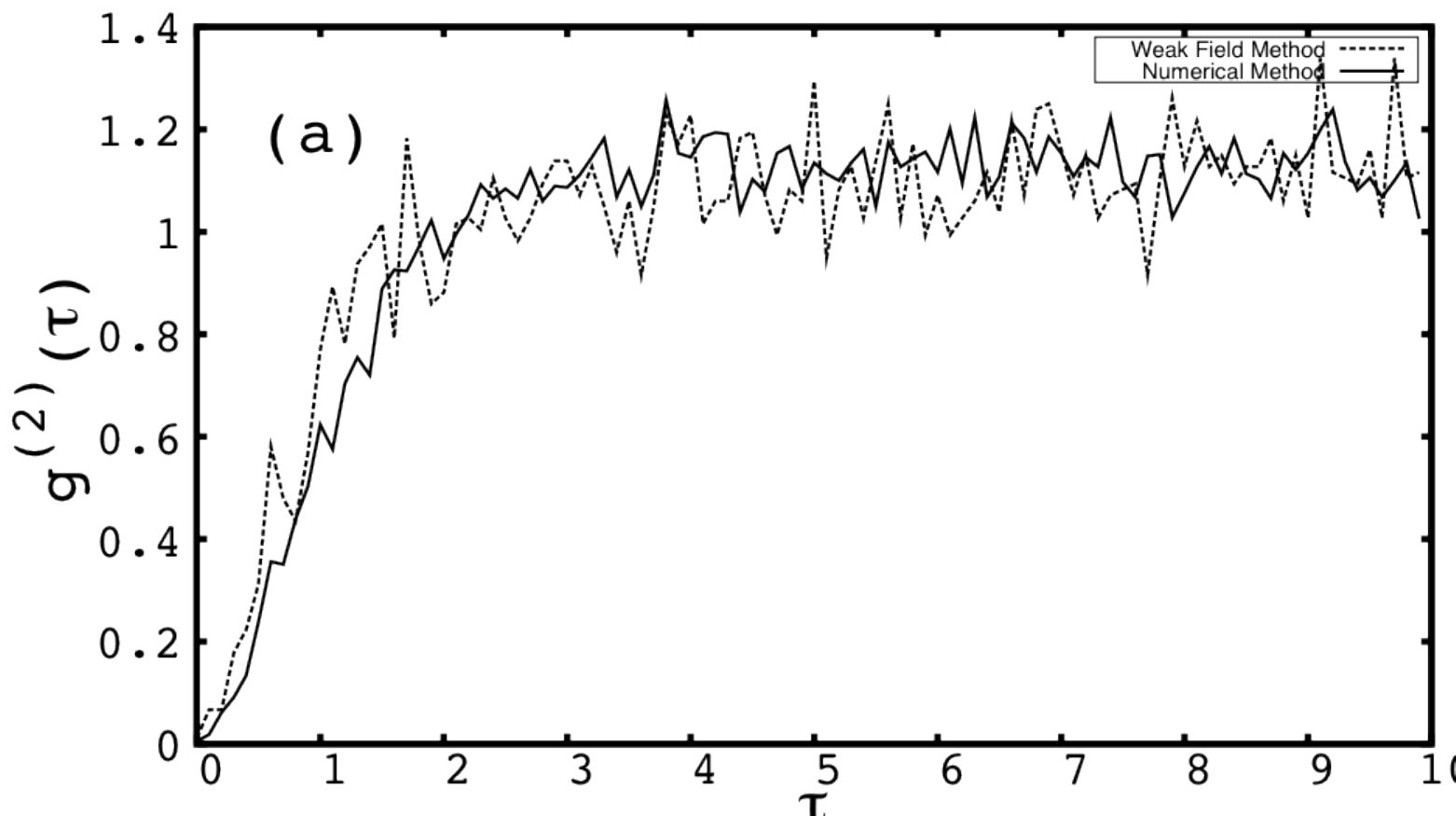}
\\\\
\includegraphics{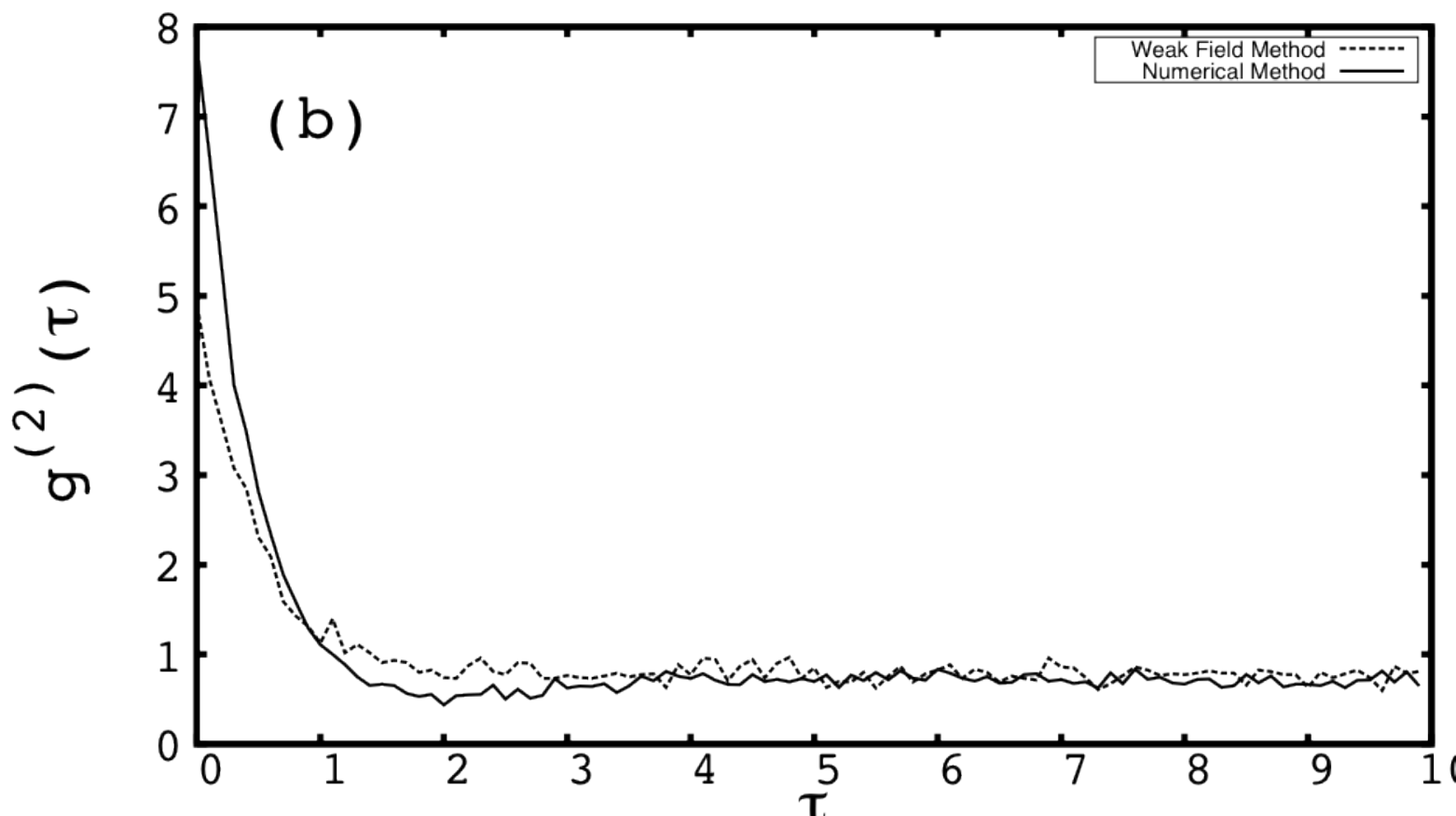}
\caption[Either site and transmission numerical correlation functions for $J=5$]{Site correlation for $J=5$ (a) shows the correlation between photons fluoresced regardless of site (b) shows the correlations between photons transmitted out the side of the cavity}
\label{fig: T=5 norm}
\end{figure}
\end{center}

The two methods agree well for all correlations except between the two sites.  There, the number of $\tau$ values are only $\sim 200$ for the weak field method.  The lack of $\tau$ values for the weak field analysis is expected since we dropped the driving term for the ground state as mention earlier.   For the other figures the number of values of $\tau$ is large enough to not be drastically affected.

The graphs of $g^{(2)}(\tau)$ for the single site fluorescence correlation, figures (\ref{fig: T=1 sites}a), (\ref{fig: T=2 sites}a), and (\ref{fig: T=5 sites}a), have the same form as the ordinary weak field trajectory model except with added oscillations.  As $J$ increases so does the frequency of oscillations.  For figure (\ref{fig: T=1 sites}a) the period is $\sim \pi$, while in figure (\ref{fig: T=2 sites}a) where $J$ doubles the period is $\sim \pi/2$.  Likewise, in figure (\ref{fig: T=5 sites}a) the period is $\sim \pi/5$.  The correlation is $T = \pi/J$ and will be derived in the next section.  The oscillations are not as clear for the correlation between site 1 and site 2.  Again, due to the low number of $\tau$ values it is difficult to determine beyond $J=1$ for the weak field method and $J=2$ for the numerical method.  However, we can see oscillations with the same period as the single site correlations.

The other figures, (\ref{fig: T=1 norm},  \ref{fig: T=2 norm}, and \ref{fig: T=5 norm}) show the correlations if we ignore the tunneling.  These agree with earlier models of a single atom in following the Hamiltonian derived in section (\ref{ss: AE}).  The results can be compared to \cite{JL} and \cite{OptComm}.

 The correlation function $g^{(2)}(\tau)$ can be rewritten in terms of the collapsed states, 
\begin{equation} 
g^{(2)}(\tau) = \frac{\braket{ \psi_c | \hat{\sigma}_+(\tau) \hat{\sigma}_-(\tau) | \psi_c}}{\braket{\hat{\sigma}_+\hat{\sigma}_-}}
\end{equation}
Where $|\psi_c\rangle =\sigma_- |\psi \rangle$ is the collapsed state.  We have seperated out $\braket{\hat{\sigma}_+\hat{\sigma}_-}$ and canceled out the same term from the bottom. Therefore, if we know which site the photon fluoresces from we know the collapsed state.  Let us assume, for now, that we are looking at $g^{(2)}(\tau)$  for the first site starting $\tau$ and stopping $\tau$.  When the first jump is observed, at $t=0$,  the wave state is collapsed to the atom being in the ground state and in the first site.  Then the quantity $\braket{ \hat{\sigma}_+ (\tau) \hat{\sigma}_-(\tau)}$ is simply the probability of being in the excited state in the first site with the initial condition of the atom starting in the ground state of the first site. 
\begin{equation}
G^{(2)}(\tau) = |C_{10}^e(\tau)|^2 = \left(\frac{Y}{\Gamma}\right)^2 \left( e^{-  \Gamma \tau} - 1 \right)^2 \mbox{cos}^2(J\tau) 
\end{equation}
We normalize to get $g^{(2)}(\tau)$ by dividing by the time average of the excited state probability which is $(Y/\sqrt{2}\Gamma)^2$.  Again this is not exactly $g^{(2)}(\tau)$ since the normalization is not exact but it is closely related.  This gives 
\begin{equation} \label{eq: g 1 to 1}
g^{(2)}_{F:1 \rightarrow 1}(\tau) =  2 \left( e^{-  \Gamma \tau} - 1 \right)^2 \mbox{cos}^2(J\tau) 
\end{equation}
The same logic can be applied to the situation when a jump from the first site starts the clock measuring $\tau$ and a jump from the second site stops the clock.  Again, the collapsed state is the atom in the ground state of the first site.  The difference now is that the unnormalized counting statistic is now proportional to the probability of being in the excited state of the second site since that is where we observe the second jump. Then
\begin{equation} \label{eq: g 1 to 2}
g^{(2)}_{F:1 \rightarrow 2}(\tau) = 2 \left( e^{-  \Gamma \tau} - 1 \right)^2 \mbox{sin}^2(J\tau) 
\end{equation}
We can also look at the correlation functions when we start the clock at the second site, which would switch the sines and cosines.  
\begin{eqnarray}
g^{(2)}_{F:2 \rightarrow 2}(\tau) &=& 2 \left( e^{-  \Gamma \tau} - 1 \right)^2 \mbox{sin}^2(J\tau)  \\
g^{(2)}_{F:2 \rightarrow 1}(\tau) &=& 2 \left( e^{-  \Gamma \tau} - 1 \right)^2 \mbox{cos}^2(J\tau) 
\end{eqnarray}
Similarly we can write the theoretical equation for $g^{(2)}(\tau)$ when we do not note which well the particle comes from.  This is equal to the probability is the sum of the probability of the atom in the excited state of either site
\begin{equation} 
g_{\gamma}^{(2)}(\tau) = |C_{10}^e(\tau)|^2 + |C_{01}^e(\tau)|^2 = \left( e^{-  \Gamma \tau} - 1 \right)^2 
\end{equation}
The first jump collapses the state to $C_{10}^g = C_{01}^g = 1/\sqrt{2}$ which are the initial conditions for the excited state.  We note that in this equation there is no longer an oscillation term dependent on the tunneling constant $J$.

To find the theoretical $g^{(2)}_{\kappa}(\tau)$ we turn to \cite{OptComm} which gives a full derivation of the function in a bad cavity, weak field limit where $g/\kappa <<1$ 
\begin{equation}
g^{(2)}_{\kappa}(\tau)= \left(1-4C^2 e^{-\frac{\gamma}{2}(1+2C)t} \right)^2
\end{equation}
 Again there is no dependence the tunneling constant $J$.

We can now combine our theoretical results with our numerical data.  We do not expect exact matches because our models have $Y=0.4$ which does not correspond to the weak field limit in which the theoretical equations were derived.  Also the models include effects from other jumps which we have not treated theoretically.  We compare the theoretical graphs with the numerical method since they give the most jumps.  For this sample $J=1$ and all other variables are the same.  We also expand the number of intervals to $2 \times 10^9$.  This now gives $\sim 10^5$  values of $\tau$ for the correlation between the sites and about four times more for the correlation between either site and the transmission graphs.   

\begin{figure}[H]
\includegraphics{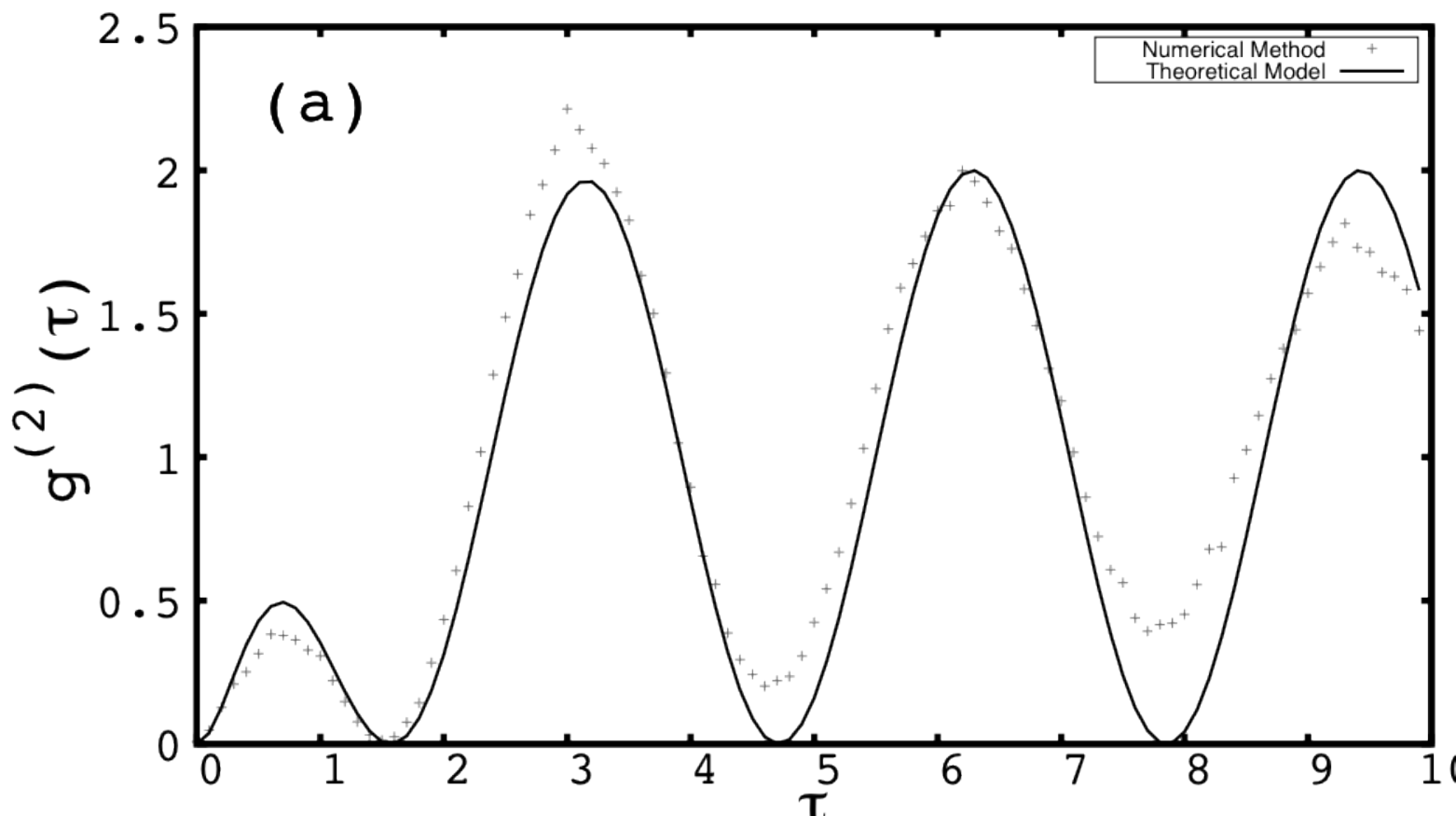}
\\\\
\includegraphics{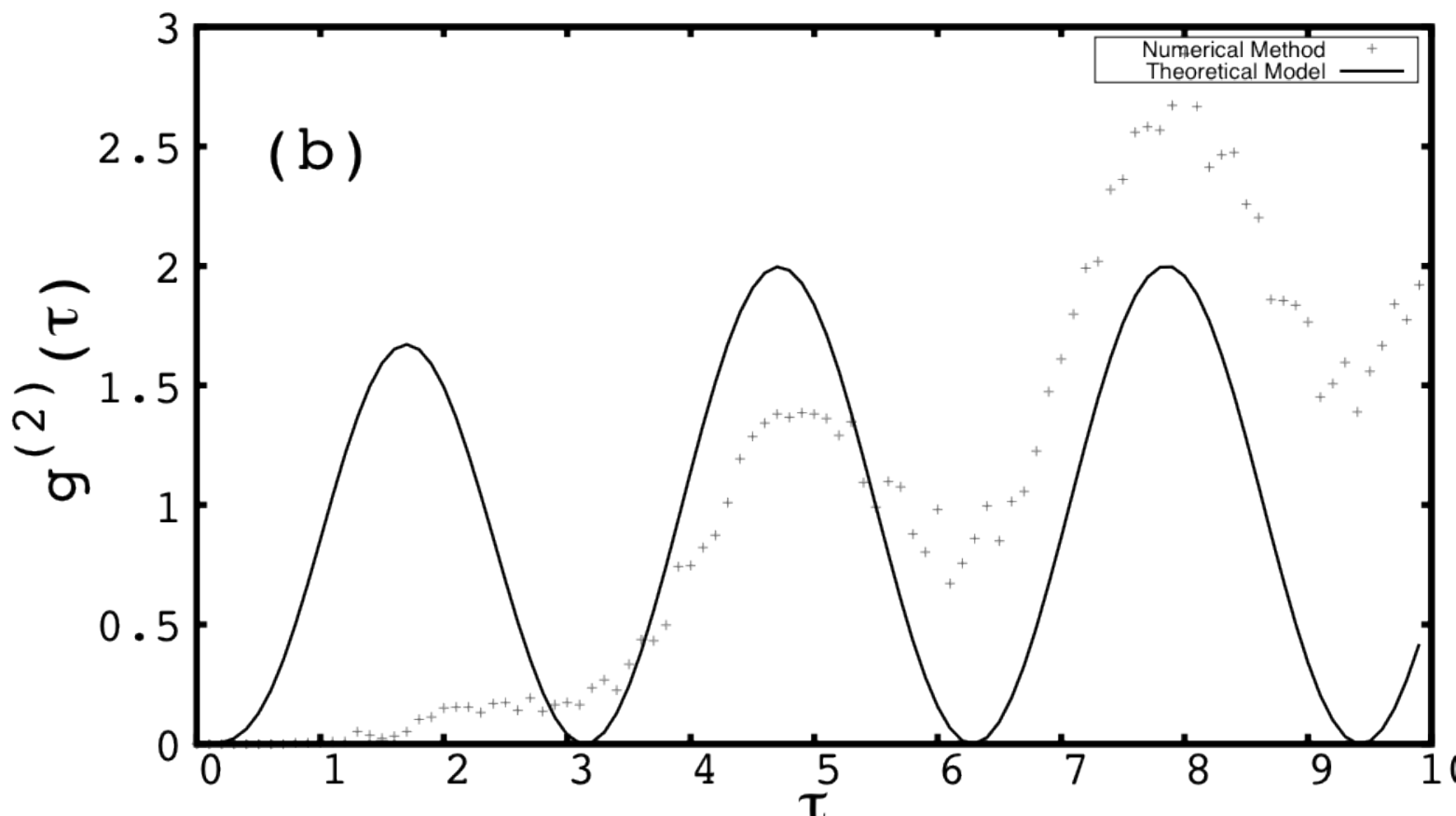}
\caption[Theoretical $g^{(2)}(\tau)$ between sites]{Comparison of theoretical and numerical correlations when $J=1$ for (a) photons fluoresced from a single cavity (b) two cavities}
\label{fig: theory comp sites}
\end{figure}

\begin{figure}[H]
\includegraphics{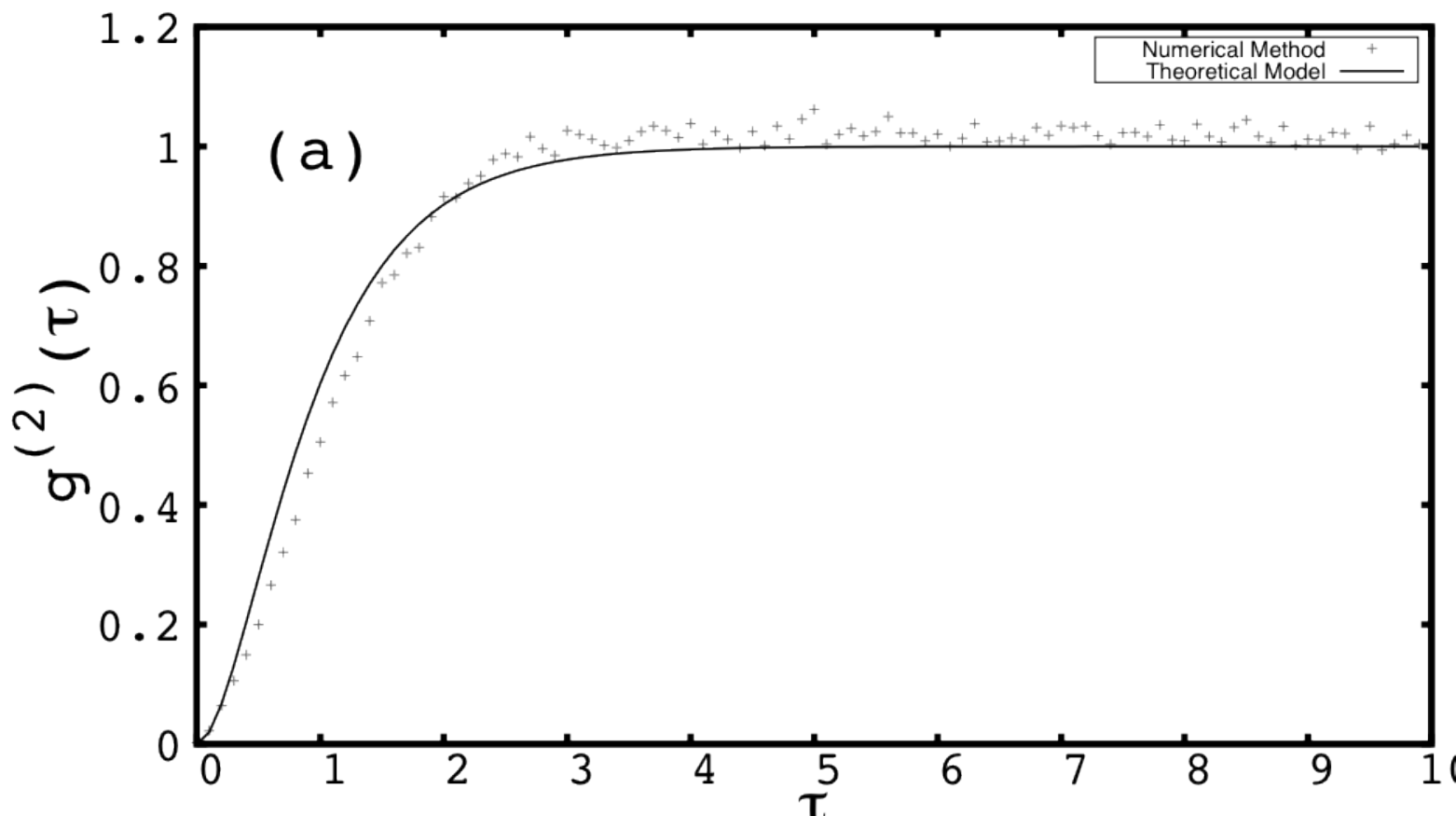}
\\\\
\includegraphics{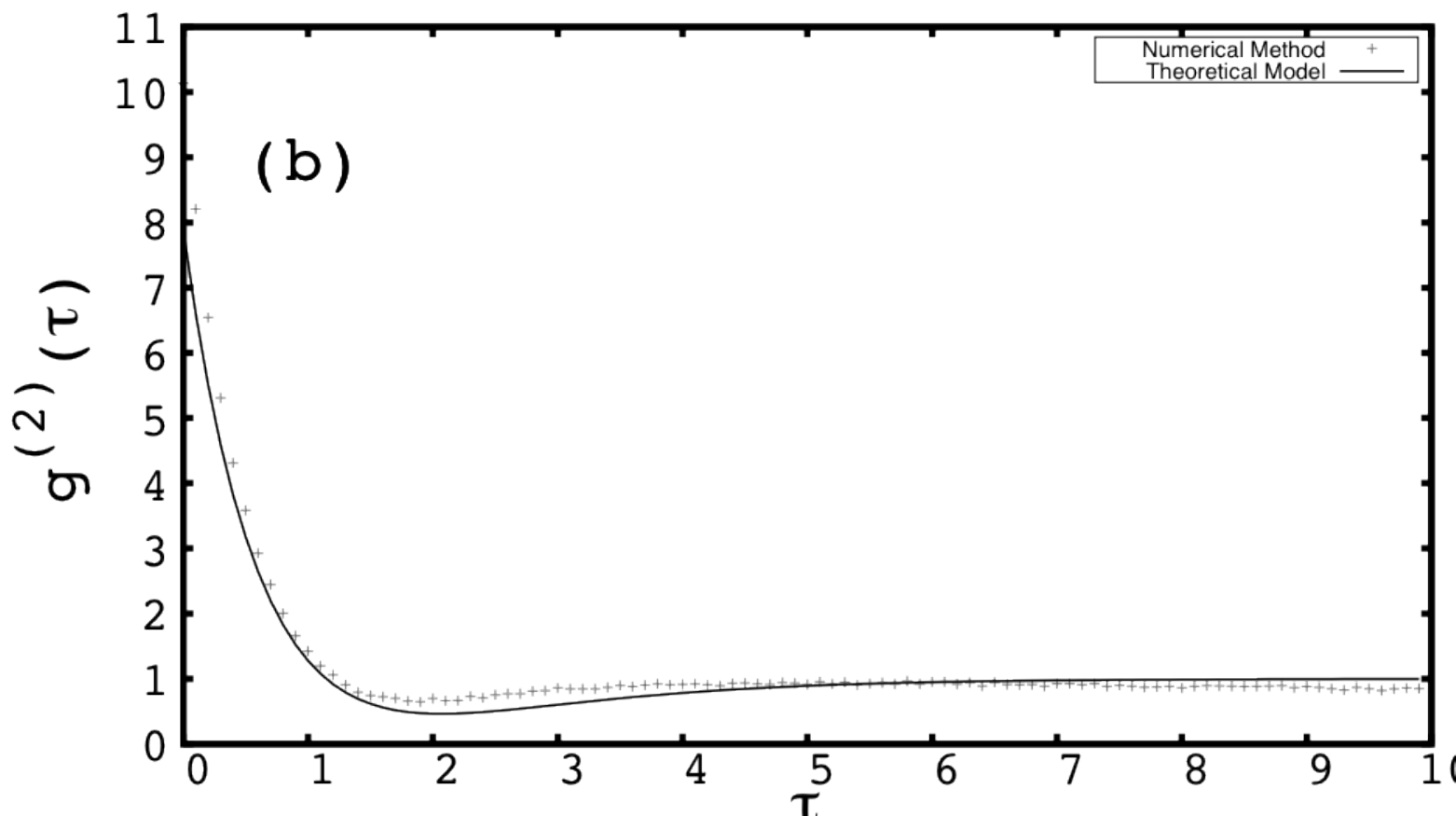}
\caption[Theoretical $g^{(2)}(\tau)$ for total fluorescence and transmission]{Comparison of theoretical and numerical correlations when $J=1$ for (a) photons fluoresced from either cavity (b) transmitted through the front of the cavity}
\label{fig: theory comp norm}
\end{figure}
\pagebreak

The theoretical model matches well in figure (\ref{fig: theory comp sites}a), however, fails to match the observed trend in (b).  For figure (\ref{fig: theory comp sites}a) we see again the period is $\sim \pi$ and now by comparison to the theoretical function of $g^{(2)}(\tau)$ we can verify that it in fact should be $\pi$ since the constant is $J=1$ and the oscillation is from the $\mbox{cos}^2$ term.  This matches our observations in the previous section of the numerical data, which showed the period was inversely dependent to the tunneling constant.  Now we can state that $\lambda = \pi/J$.  The osillations in the model begin to dissipate for higher $\tau$ due to other collapses inbetween making longer correlation times less likely to oscillate.  The same oscillations are seen in figure  (\ref{fig: theory comp sites}b).  These do not follow the same trend as the previous graph for reasons we do not fully understand.  A likely explaination is related to having less values of $\tau$ meaning that there may be more jumps inbetween the collapse the states.

The other correlations, figure (\ref{fig: theory comp norm}), also roughly follow our theoretical expectations.  The normalization is altered to reflect since for $\tau<2$ the value of $g^{(2)}(\tau$ varries from $\tau > 2$ which is the steady state solution.  Therefore, we ignore the $\tau$ counts $<2$ and divide by the counts for $\tau>2$.  For figure  (\ref{fig: theory comp norm}a) the theory matches with some small deviations around $\tau=1$ due to the new normalization scheme.  Figure (\ref{fig: theory comp norm}b) matches except for very small values of $\tau$ again due to the new normalization. Both match much better because there on the order of 10 more values of $\tau$ than the single site correlations.

\subsection{ Conclusions}
Theoretical solutions of $g^{(2)}(\tau)$ and the numerical models agree that the correlation between photon fluorescence when measured between sites is proportional to the tunneling rate.  Looking at the problem qualitatively this agrees with what we would expect.  The wave function will only be collapsed to that site with a jump for a set amount of time.  After that, it will collapse to the other site.  If we say the atom emits a photon from site $a$ at $t=0$ then the wave function of the atom is collapsed to the ground state in $a$.  The atom, however, will not stay in $a$ much longer since its wave function will now evolve making it more likely to be in the other site $b$.  Now, there is a chance the atom will experience a collapse again, at time $t=\tau$, in $b$ but it could also move back to $a$.  During this time the probability of detecting a jump from site $a$ is zero or very small.  However, once the atom tunnels back to $a$ the probability increases.  This is why there is an oscillation in $g^{(2)}(\tau)$.  After the first collapse at $t=0$ the atom will leave $a$ (small values of $\tau$) meaning there is little chance of a second jump.  The atom can travel back and forth many times before the second jump.  If we are looking at the correlation between photon counts from a single well then when the atom is in site $b$ the probability of a count is very small.  Therefore, the oscillation is driven by the tunneling between sites since there are no counts possible at times when the atom is in the other site.  

The other plots, of the correlation between photons fluoresced from either site and the correlation of photons transmitted, verify the results of the trajectory model.  If we do not note which site the fluoresced photons come from we are ignoring the optical lattice and the tunneling term in the Hamiltonian which leaves us with only the trajectory terms that have been previously investigated in \cite{OptComm} and \cite{CAVQEDBIG}.  The lattice does not affect the transmitted photons since the lattice is created on the same axis that the transmitted photons are emitted from.  Therefore, we cannot tell which site the photon comes.  Neither of the theoretical or numerical plots show any dependence on the tunneling constant $J$

These tests show that in fact our theoretical and numerical models do follow the expected trends and verify that $g^{(2)}(\tau)$ can be used to determine the tunneling constant.  Since we know that the model follows the behavior predicted earlier we can continue to expand the model by adding more atoms making the system into a BEC.

We verified the dependence of $g^{(2)}(\tau)$ on $J$ for the single site fluorescence correlations and independence of $J$ for either site and transmission correlations.  The numerical models we made match our theoretical values for the correlation function for a single site, however, diverge when looking at the correlation between two sites.  They also are not exactly match when looking at fluorescence from either site or transmission.  Some of these patterns can be explained with our choice of $g$, $\kappa,$ and $\gamma$ especially with the transmission correlation.  However, there may need to be a better method developed to normalize the jump counts.  This is likely causing the difference in the correlation from either site and likely between two sites.  Larger sample sizes may also help since after $\tau=20$ the counting statistic becomes prone to noise from the random number generator.  Larger sample sizes would reduce the fraction of the noise.

We have also established methods to model more complicated multi-site and multi-atom systems numerically.  A larger simulation with three sites and two particles was run but since there are more states (12 with the bad cavity limit, 36 without) it takes much longer, see Appendix A.  In the plots developed there were not recognizable oscillations dependent on $J$.  Further work is needed to speed up the programs to find the jump times.  It is also unlikely that a theoretical solution for $g^{(2)}(\tau)$ can be found since the dimensions of multi-site and multi-atom models can grow quickly.  Therefore, future work will likely be dependent on numerical results.

\section{Appendix A: Amplitude rates for three site, two atom setup}
For the notation, ${}^n C_{xyz}^{g,ie}$, $n$ is the number of photons, $x,y,z$ are the number of atoms in each site, and $g, ie$ are the number of atoms in the excited state where $g=0$ atoms excited, $i = 1,2$ atoms excited.

Ground state
\begin{eqnarray}
^0 \dot{C}_{101}^g &=& i J ( {}^0 C_{110}^g + {}^0 C_{011}^g) \\
^0 \dot{C}_{110}^g &=& i J ({}^0 C_{101}^g + \sqrt{2} {}^0 C^g_{200} + \sqrt{2} {}^0 C_{020}^g) \\
^0 \dot{C}_{011}^g &=& i J ({}^0 C_{110}^g + \sqrt{2} {}^0 C^g_{002} + \sqrt{2} {}^0 C_{020}^g) \\
^0 \dot{C}_{200}^g &=& i J {}^0 C_{110}^g - i \hbar U {}^0 C_{200}^g \\
^0 \dot{C}_{020}^g &=& i J \sqrt{2}({}^0 C_{110}^g + {}^0 C_{011}^g) - i \hbar U {}^0 C_{020}^g \\
^0 \dot{C}_{002}^g &=& i J \sqrt{2} {}^0 C_{011}^g - i \hbar U {}^0 C_{002}^g
\end{eqnarray}
1st excited states
\begin{eqnarray}
^1 \dot{C}_{101}^g &=& i J ( {}^1 C_{110}^g + {}^1 C_{011}^g) + Y {}^0 C_{101}^g + g \sqrt{2} {}^0 C_{101}^{1e} - \kappa {}^1 C_{101}^g\\
^1 \dot{C}_{110}^g &=& i J ({}^1 C_{101}^g + \sqrt{2} {}^1 C^g_{200} + \sqrt{2} {}^1 C_{020}^g) + Y {}^0 C_{110}^g + g \sqrt{2} {}^0 C_{110}^{1e} - \kappa {}^1 C_{110}^g\\
^1 \dot{C}_{011}^g &=& i J ({}^1 C_{110}^g + \sqrt{2} {}^1 C^g_{002} + \sqrt{2} {}^1 C_{020}^g) + Y {}^0 C_{011}^g + g \sqrt{2} {}^0 C_{011}^{1e} - \kappa {}^1 C_{011}^g\\
^1 \dot{C}_{200}^g &=& i J {}^1 C_{110}^g - i \hbar U {}^1 C_{200}^g + Y {}^0 C_{200}^g + g \sqrt{2} {}^0 C_{200}^{1e} - \kappa {}^1 C_{200}^g\\
^1 \dot{C}_{020}^g &=& i J \sqrt{2}({}^1 C_{110}^g + {}^1C_{011}^g) - i \hbar U {}^1 C_{020}^g + Y {}^0 C_{020}^g + g \sqrt{2} {}^0 C_{020}^{1e} - \kappa {}^1 C_{020}^g\\
^1 \dot{C}_{002}^g &=& i J \sqrt{2} {}^1 C_{011}^g - i \hbar U {}^1 C_{002}^g+ Y {}^0 C_{002}^g + g \sqrt{2} {}^0 C_{002}^{1e} - \kappa {}^1 C_{002}^g\\
\nonumber \\
^0 \dot{C}_{101}^{1e} &=& i J ( {}^0 C_{110}^{1e} + {}^0 C_{011}^{1e}) - g \sqrt{2} {}^1 C_{101}^{g} - \frac{\gamma}{2} {}^0 C_{101}^{1e}\\
^0 \dot{C}_{110}^{1e} &=& i J ({}^0 C_{101}^{1e} + \sqrt{2} {}^0 C^{1e}_{200} + \sqrt{2} {}^0 C_{020}^{1e}) - g \sqrt{2} {}^1 C_{110}^{g} - \frac{\gamma}{2} {}^0 C_{110}^{1e}\\
^0 \dot{C}_{011}^g &=& i J ({}^0 C_{110}^{1e} + \sqrt{2} {}^0 C^{1e}_{002} + \sqrt{2} {}^0 C_{020}^{1e}) - g \sqrt{2} {}^1 C_{011}^{1e} -\frac{\gamma}{2} {}^0 C_{011}^{1e}\\
^0 \dot{C}_{200}^{1e} &=& i J {}^0 C_{110}^{1e} - i \hbar U {}^0 C_{200}^{1e} - g \sqrt{2} {}^1 C_{200}^{g} - \frac{\gamma}{2} {}^0 C_{200}^{1e}\\
^0 \dot{C}_{020}^{1e} &=& i J \sqrt{2}({}^0 C_{110}^{1e} + {}^0 C_{011}^{1e}) - i \hbar U {}^0 C_{020}^{1e} - g \sqrt{2} {}^1 C_{020}^{g} - \frac{\gamma}{2} {}^0 C_{020}^{1e}\\
^0 \dot{C}_{002}^{1e} &=& i J \sqrt{2} {}^0 C_{011}^{1e} - i \hbar U {}^0 C_{002}^{1e} - g \sqrt{2} {}^1 C_{002}^{g} -\frac{\gamma}{2} {}^0 C_{002}^{1e}\\
\end{eqnarray}

2nd excited states
\begin{eqnarray}
^2 \dot{C}_{101}^g &=& i J ( {}^2 C_{110}^g + {}^2 C_{011}^g) + Y \sqrt{2} {}^1 C_{101}^g + 2 g  {}^1 C_{101}^{1e} - 2\kappa {}^2 C_{101}^g\\
^2 \dot{C}_{110}^g &=& i J ({}^2 C_{101}^g + \sqrt{2} {}^2 C^g_{200} + \sqrt{2} {}^2 C_{020}^g) + Y \sqrt{2} {}^1 C_{110}^g + 2 g  {}^1 C_{110}^{1e} \nonumber \\
&& - 2\kappa {}^2 C_{110}^g\\
^2 \dot{C}_{011}^g &=& i J ({}^2 C_{110}^g + \sqrt{2} {}^2 C^g_{002} + \sqrt{2} {}^2 C_{020}^g) + Y \sqrt{2} {}^1 C_{011}^g + 2 g {}^1 C_{011}^{1e} \nonumber \\
&& - 2\kappa {}^2 C_{011}^g\\
^2 \dot{C}_{200}^g &=& i J {}^2 C_{110}^g - i \hbar U {}^2 C_{200}^g + Y \sqrt{2} {}^1 C_{200}^g + 2 g {}^1 C_{200}^{1e} - 2\kappa {}^2 C_{200}^g\\
^2 \dot{C}_{020}^g &=& i J \sqrt{2}({}^2 C_{110}^g + {}^2 C_{011}^g) - i \hbar U {}^2 C_{020}^g + Y \sqrt{2}{}^1 C_{020}^g + 2 g {}^1 C_{020}^{1e} \nonumber \\
&& - 2\kappa {}^2 C_{020}^g\\
^2 \dot{C}_{002}^g &=& i J \sqrt{2} {}^2 C_{011}^g - i \hbar U {}^2 C_{002}^g+ Y\sqrt{2} {}^1 C_{002}^g +  2 g  {}^1 C_{002}^{1e} - 2\kappa {}^2 C_{002}^g \\
\nonumber \\
^1 \dot{C}_{101}^{1e} &=& i J ( {}^1 C_{110}^{1e} + {}^1 C_{011}^{1e}) + Y  {}^2 C_{101}^{g} -  2 g {}^2 C_{101}^{g} - \kappa {}^1 C_{101}^{1e} - \frac{\gamma}{2} {}^1 C_{101}^{1e}\\
^1 \dot{C}_{110}^{1e} &=& i J ({}^1 C_{101}^{1e} + \sqrt{2} {}^1 C^{1e}_{200} + \sqrt{2} {}^1 C_{020}^{1e}) + Y  {}^2 C_{110}^{g} -  2 g {}^2 C_{110}^{g}  \nonumber \\
&& - \kappa {}^1 C_{110}^{1e} - \frac{\gamma}{2} {}^1 C_{110}^{1e}\\
^1 \dot{C}_{011}^{1e} &=& i J ({}^1 C_{110}^{1e} + \sqrt{2} {}^1 C^{1e}_{002} + \sqrt{2} {}^1 C_{020}^{1e}) + Y  {}^2 C_{011}^{g} -  2 g {}^2 C_{011}^{g}\nonumber \\
&&  - \kappa {}^1 C_{011}^{1e} - \frac{\gamma}{2} {}^1 C_{011}^{1e}\\
^1 \dot{C}_{200}^{1e} &=& i J {}^1 C_{110}^{1e} - i \hbar U {}^1 C_{200}^{1e} + Y  {}^2 C_{200}^{g} -  2 g {}^2 C_{200}^{g} - \kappa {}^1 C_{200}^{1e} - \frac{\gamma}{2} {}^1 C_{200}^{1e}\\
^1 \dot{C}_{020}^{1e} &=& i J \sqrt{2}({}^1 C_{110}^{1e}+ {}^1 C_{011}^{1e}) - i \hbar U {}^1 C_{020}^{1e}  + Y  {}^2 C_{020}^{g} -  2 g {}^2 C_{020}^{g} \nonumber \\
&& - \kappa {}^1 C_{020}^{1e} - \frac{\gamma}{2} {}^1 C_{020}^{1e}\\
^1 \dot{C}_{002}^{1e} &=& i J \sqrt{2} {}^1 C_{011}^{1e} - i \hbar U {}^1 C_{002}^{1e} + Y  {}^2 C_{002}^{g} -  2 g {}^2 C_{002}^{g} - \kappa {}^1 C_{002}^{1e} - \frac{\gamma}{2} {}^1 C_{002}^{1e}\\
\nonumber \\
^0 \dot{C}_{101}^{2e} &=& i J ( {}^0 C_{110}^{2e} + {}^0 C_{011}^{2e}) - g \sqrt{2} {}^1 C_{101}^{1e} - \gamma {}^0 C_{101}^{2e}\\
^0 \dot{C}_{110}^{2e} &=& i J ({}^0 C_{101}^{2e} + \sqrt{2} {}^0 C^{2e}_{200} + \sqrt{2} {}^0 C_{020}^{1e}) - g \sqrt{2} {}^1 C_{110}^{1e} - \gamma {}^0 C_{110}^{2e}\\
^0 \dot{C}_{011}^{2e} &=& i J ({}^0 C_{110}^{2e} + \sqrt{2} {}^0 C^{2e}_{002} + \sqrt{2} {}^0 C_{020}^{1e}) - g \sqrt{2} {}^1 C_{011}^{1e} -\gamma {}^0 C_{011}^{2e}\\
^0 \dot{C}_{200}^{2e} &=& i J {}^0 C_{110}^{2e} - i \hbar U {}^0 C_{200}^{2e} - g \sqrt{2} {}^1 C_{200}^{1e} - \gamma {}^0 C_{200}^{2e}\\
^0 \dot{C}_{020}^{2e} &=& i J \sqrt{2}({}^0 C_{110}^{2e} + {}^0 C_{011}^{2e}) - i \hbar U {}^0 C_{020}^{2e} - g \sqrt{2} {}^1 C_{020}^{1e} - \gamma {}^0 C_{020}^{2e}\\
^0 \dot{C}_{002}^{2e} &=& i J \sqrt{2} {}^0 C_{011}^{2e} - i \hbar U {}^0 C_{002}^{2e} - g \sqrt{2} {}^1 C_{002}^{1e} -\gamma {}^0 C_{002}^{2e}
\end{eqnarray}

\end{document}